\documentclass[acmsmall]{acmart}

\usepackage{algorithm}
\usepackage{algorithmicx}
\usepackage{algpseudocode}
\usepackage{amsmath}
\usepackage{graphicx}
\usepackage{balance}
\usepackage{cleveref}
\usepackage{xcolor}
\usepackage{wrapfig}
\usepackage{xspace}
\usepackage{subcaption}
\usepackage{listings}
\usepackage{tcolorbox}
    \tcbuselibrary{skins}
\usepackage{tabularx}
\usepackage{booktabs}
\usepackage{colortbl}
\usepackage{multirow}

\setcopyright{cc}
\setcctype{by}

\algnewcommand\algorithmicforeach{\textbf{for each}}
\algdef{S}[FOR]{ForEach}[1]{\algorithmicforeach\ #1\ \algorithmicdo}

\algnewcommand\algorithmicinput{\textbf{Input:}}
\algnewcommand\Input{\item[\algorithmicinput]}
\algnewcommand\algorithmicoutput{\textbf{Output:}}
\algnewcommand\Output{\item[\algorithmicoutput]}
\algrenewcommand\algorithmiccomment[1]{\State\textcolor{gray}{// #1}}

\newcommand{\Continue}{\textbf{continue}}

\algdef{SE}[REPEATN]{RepeatN}{EndRepeatN}[1]{\algorithmicrepeat\ #1 \textbf{times}}{\algorithmicend}

\newcommand{\crossmark}{$\times$}
\newcommand{\FRAM}{SeqFuzzSDN\xspace}
\newcommand{\FRAMns}{SeqFuzzSDN$^{NS}$\xspace}

\newcommand{\DELTA}{\textsc{Delta}\xspace}
\newcommand{\DELTAe}{$\textsc{Delta}^E$\xspace}
\newcommand{\BEADS}{\textsc{Beads}\xspace}
\newcommand{\BEADSe}{$\textsc{Beads}^E$\xspace}
\newcommand{\FUZZSDN}{\textsc{FuzzSDN}\xspace}
\newcommand{\FUZZSDNe}{$\textsc{FuzzSDN}^E$\xspace}

\newcommand*\card[1]{\lvert#1\rvert}

\newcolumntype{Y}{>{\centering\arraybackslash}X}

\title{Learning-Guided Fuzzing for Testing Stateful SDN Controllers}

\author{Rapha{\"e}l Ollando}
\email{raphael.ollando@uni.lu}
\affiliation{\institution{University of Luxembourg}
	\streetaddress{29 Avenue John F. Kennedy}
	\city{Luxembourg}
	\postcode{1859}
	\country{Luxembourg}
}

\author{Seung Yeob Shin}
\email{seungyeob.shin@uni.lu}
\affiliation{\institution{University of Luxembourg}
	\streetaddress{29 Avenue John F. Kennedy}
	\city{Luxembourg}
	\postcode{1859}
	\country{Luxembourg}
}

\author{Lionel C. Briand}
\authornote{Part of this work was done when he was affiliated with the Interdisciplinary Centre for Security, Reliability, and Trust (SnT) of the University of Luxembourg.}
\email{lionel.briand@lero.ie}
\affiliation{\institution{Lero SFI Centre for Software Research, University of Limerick}
	\streetaddress{Tierney building}
	\city{Limerick}
	\postcode{V94 NYD3}  
	\country{Ireland}
}
\affiliation{\institution{University of Ottawa}
	\streetaddress{800 King Edward Avenue}
	\city{Ottawa}
	\postcode{ON K1N 6N5}  
	\country{Canada}
}

\begin{document}

\begin{abstract}
Controllers for software-defined networks (SDNs) are centralised software components that enable advanced network functionalities, such as dynamic traffic engineering and network virtualisation. 
However, these functionalities increase the complexity of SDN controllers, making thorough testing crucial.
SDN controllers are stateful, interacting with multiple network devices through sequences of control messages.
Identifying stateful failures in an SDN controller is challenging due to the infinite possible sequences of control messages, which result in an unbounded number of stateful interactions between the controller and network devices.
In this article, we propose \FRAM, a learning-guided fuzzing method for testing stateful SDN controllers.
\FRAM aims to (1)~efficiently explore the state space of the SDN controller under test, (2)~generate effective and diverse tests (i.e., control message sequences) to uncover failures, and (3)~infer accurate failure-inducing models that characterise the message sequences leading to failures.
In addition, we compare \FRAM with three extensions of state-of-the-art (SOTA) methods for fuzzing SDNs.
Our findings show that, compared to the extended SOTA methods, \FRAM (1)~generates more diverse message sequences that lead to failures within the same time budget, and (2)~produces more accurate failure-inducing models, significantly outperforming the other extended SOTA methods in terms of sensitivity. 
\end{abstract}

\ccsdesc[500]{Networks~Programmable networks}
\ccsdesc[500]{Software and its engineering~Software testing and debugging}

\keywords{Software-Defined Networks, Software Testing, Fuzzing}

\maketitle

\section{Introduction}
\label{sec:intro}

Software-defined networks (SDNs)~\cite{SDN:15}, which have been applied in many domains such as data centres~\cite{Drutskoy2013, Wang2017}, satellite communications~\cite{Ferrus2016,LiuSZCSK18}, and the Internet of Things~\cite{Rafique2020,Shin2020:SEAMS}, have gained popularity due to the programmability of their controllers, enabling the deployment of network services through software.
An SDN controller is a centralised software component in the SDN that enables the implementation of advanced network functionalities, such as dynamic traffic engineering~\cite{Shin2020:SEAMS} and network virtualisation~\cite{BlenkBRK2016}.
However, implementing such functionalities increases the complexity of SDN controllers.
Furthermore, having a centralised controller introduces new attack surfaces (e.g., ARP spoofing for SDN~\cite{Wagner2001,AlharbiDPP2016}) that can be exploited by malicious users to manipulate SDNs. 
Hence, testing SDN controllers becomes even more important and poses specific challenges compared to testing traditional networks, which typically lack software controllers and provide static network operations.

An SDN controller is a stateful software component that maintains a holistic view of the SDN, capturing the state information of network devices, links, and the controller itself.
This enables the controller to provide dynamic network operations in an efficient and effective manner.
However, testing stateful SDN controllers is challenging. 
An SDN controller interacts with multiple network devices through sequences of inbound and outbound control messages defined in the underlying SDN communication protocol (e.g., OpenFlow~\cite{OpenFlowSpec}).
If a failure can occur only in a certain state of an SDN controller, discovering such a stateful failure requires engineers to identify message sequences that bring the controller into that state.
However, discovering such stateful failures is a hard problem due to the potentially infinite number of possible sequences of control messages.
This is because the size of sequences is unbounded, and there are various types of control messages with different sizes.
In addition, even if engineers obtain sequences of control messages that cause failures, manual inspection of these sequences is time-consuming and error-prone.
This may result in misunderstandings of the causes of failures and hence the application of unreliable fixes.

Fuzzing techniques have been widely applied for testing various network systems~\cite{Natella22,PhamBR2020:AFLNET,Lee2017:DELTA,Jero2017:BEADS}.
Among these, state-aware fuzzing techniques that do not depend on protocol specifications could be considered for testing SDN controllers, as, to our knowledge, no existing state-aware fuzzing techniques account for the specificities (e.g., architecture and protocol) of SDNs.
For example, AFLNet~\cite{PhamBR2020:AFLNET} constructs finite state machines (FSMs) based on the response codes generated by the server under test and uses these FSMs to guide the fuzzing process.
AFLNet employs common byte-level fuzz operators, such as bit flipping as well as the insertion, deletion, and substitution of byte blocks.
However, AFLNet operates under the working assumption that communication protocols embed special codes in response messages, which is not always the case, as in our SDN context.
StateAFL~\cite{Natella22} infers FSMs based on the in-memory states of the server, leveraging compile-time instrumentation and fuzzy hashing techniques; hence, it does not require response codes.
During the fuzzing process, StateAFL guides the generation of new inputs to the server based on the inferred FSMs.
It employs both byte-level and message-level fuzz operators, which do not rely on protocol specifications.
NSFuzz~\cite{Qin2023:TOSEM} uses a combination of static analysis and manual annotation on the server's source code to identify states based on program variables and construct FSMs that capture the transitions between these states.
It then performs FSM-guided fuzzing using fuzz operators similar to those in AFLNet.
However, the state-aware fuzzing techniques introduced in this research strand are applicable to the server-client architecture by replacing a client with a fuzzer.
The fuzzer replays captured message sequences and modifies them during the fuzzing process.
In contrast, the SDN architecture differs significantly from the server-client architecture.
For example, in the SDN architecture, communication is initiated between an SDN controller and switches, whereas in the server-client architecture, clients typically initiate requests to the server.
Additionally, SDN switches also communicate with one another to enable network communication and services.
Therefore, replacing an SDN switch with a fuzzer for testing an SDN controller is challenging. Furthermore, the working assumptions of these techniques, such as response codes, compile-time instrumentation, and source-code analysis and annotation, make them difficult to apply when testing an SDN controller.
SDN operators are more concerned with potential failures that can occur in realistic scenarios, such as when a malicious user intercepts messages and disrupts the SDN during its operation~\cite{Lee2017:DELTA,Jero2017:BEADS,OllandoSB2023:FuzzSDN,Canini2012:NICE,Dhawan2015:SPHINX}.

There are some prior studies~\cite{Lee2017:DELTA,Jero2017:BEADS,OllandoSB2023:FuzzSDN} that test SDN controllers by taking into account the architecture and protocols of SDNs.
For example, \DELTA~\cite{Lee2017:DELTA} is a security assessment framework for SDNs.
It reproduces existing SDN-related attack scenarios and uncovers new ones through fuzzing.
Specifically, in fuzzing, \DELTA modifies control messages by treating them as byte streams and randomising them.
\BEADS~\cite{Jero2017:BEADS} is an automated attack discovery tool for SDNs.
In contrast to \DELTA, \BEADS fuzzes control messages while adhering to the SDN protocol (i.e., OpenFlow), aiming to create test scenarios that can exercise components beyond the protocol parsers of SDN controllers.
FuzzSDN~\cite{OllandoSB2023:FuzzSDN} also adheres to the SDN protocol in its fuzzing process to test components beyond the protocol parsers of an SDN controller.
In addition, FuzzSDN employs machine learning techniques to infer failure-inducing models that characterise the conditions under which failures occur, and uses them to guide the fuzzing.
These techniques position their fuzzers between the SDN controller and the SDN switches to sniff and modify control messages, leveraging the man-in-the-middle attack strategy~\cite{ContiDL2016}.
Hence, they do not require any modifications, replacements, annotations, or instrumentation of the components (i.e., switches and controllers) in SDNs, enabling the testing of SDN controllers in a realistic setting.
However, these techniques, which account for the architecture and protocols of SDNs, do not consider the stateful nature of SDN controllers.

\textbf{Contributions.}
In this article, we propose \FRAM, a learning-guided fuzzing method for testing stateful SDN controllers.
\FRAM aligns with the aforementioned research strand that leverages the architecture and protocols of SDNs.
Hence, \FRAM tests SDN controllers in a realistic operational setting without requiring any compile-time instrumentation, manual annotation of source code, and replacing an SDN switch with a fuzzer.
Instead, \FRAM sniffs and fuzzes control messages exchanged between the SDN controller and switches by being aware of the stateful behaviours of the controller.
\FRAM employs a fuzzing strategy guided by Extended Finite State machines (EFSMs) in order to
(1)~efficiently explore the space of states of the SDN controller under test,
(2)~generate effective and diverse tests (i.e., message sequences) to uncover failures, and
(3)~infer accurate EFSMs that characterise the sequences of control messages leading to failures.
Note that since the SDN communication protocol specifies various message fields, their values, and relations, guard conditions on state transitions in EFSMs are well-suited to capture state changes associated with such message fields, values, and relations.

We evaluated \FRAM by applying it to two well-known open-source SDN controllers: ONOS~\cite{Berde2014:ONOS} and RYU~\cite{RYU}.
Additionally, we compared \FRAM against our extensions of three state-of-the-art (SOTA) methods---\DELTA~\cite{Lee2017:DELTA}, \BEADS~\cite{Jero2017:BEADS}, and \FUZZSDN~\cite{OllandoSB2023:FuzzSDN}---which were used as baselines for generating tests for SDN controllers. 
We extended \DELTA, \BEADS, and \FUZZSDN to produce EFSM models, since these SOTA methods were not originally designed to generate such models.
It is important to note that, these three baselines are the best available options for evaluating \FRAM when testing SDN controllers by fuzzing control messages.
Our experiment results show that \FRAM significantly outperforms the three baselines.
Specifically, compared to the baselines, \FRAM generates more diverse and effective tests (i.e., message sequences) that lead to failures, as well as more accurate EFSMs that characterise failure-inducing message sequences. 
In addition, \FRAM can be applied to large SDNs since its performance is independent of the network size.
Our complete evaluation results and the \FRAM tool can be accessed online~\cite{Artifacts}.

\textbf{Organisation.}
The remainder of this article is structured as follows:
Section~\ref{sec:background and problem} provides the background and defines the problem of testing stateful SDN controllers.
Section~\ref{sec:approach} details the steps of \FRAM.
Section~\ref{sec:evaluation} presents the empirical evaluation of \FRAM.
Section~\ref{sec:related works} compares \FRAM with related work.
Finally, Section~\ref{sec:conclusion} concludes the article.

\section{Background and problem description}
\label{sec:background and problem}

This section introduces SDN concepts, including the SDN architecture and control messages. 
We then define the problem of testing stateful SDN controllers.

\textbf{Architecture.}
The SDN architecture~\cite{SDN:15} separates the network into the control plane and the data plane, which is a key distinction from traditional networks that do not possess this separation.
In the control plane, an SDN controller provides network administrators with a global view of the network, enabling centralised control over network operations.
The centralised control allows administrators to optimally manage network resources and to effectively enforce network policies. 
Furthermore, an SDN controller provides engineers with APIs~\cite{SDN:15}, enabling them to develop and install custom applications on the controller to address system-specific needs (e.g., dynamic adaptive traffic control application~\cite{Shin2020:SEAMS}).
The data plane consists of SDN switches, which are responsible for forwarding data messages (i.e., data packets) based on the instructions provided by the controller.
SDN switches are connected to hosts that generate and receive data messages.
Such hosts can be servers, clients, and IoT devices in a networked system. 
In the SDN architecture, the SDN controller serves as the central software component that enables the provision of flexible and efficient network services.

\begin{figure}[t]
    \centering
    \includegraphics[width=0.6\linewidth]{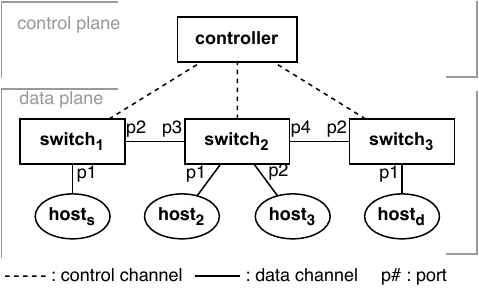}
    \caption{An SDN topology example.}
    \label{fig:SDN example}
    \Description{}
\end{figure}

Figure~\ref{fig:SDN example} depicts an SDN topology example that consists of a controller, three switches, and four hosts.
The controller communicates with the three switches via control channels that carry control messages (e.g., OpenFlow messages~\cite{OpenFlowSpec}).
The switches and hosts, on the other hand, are connected via data channels that carry data messages encapsulated by standard network protocols such as ARP~\cite{Plummer1982:ARP, PetersonD2007:CompNet} and IP~\cite{Postel1981:IP, PetersonD2007:CompNet}.

\begin{table*}[t]
\caption{An example sequence of messages for discovering host locations. The messages in this table are generated by the hosts, switches, and a controller depicted in Figure~\ref{fig:SDN example}.}
\label{tbl:message sequence}
\centering
\small
\begin{tabularx}{\textwidth}{l X Y Y Y}
\toprule
$m_{i}$ & message & sender & receiver & channel \\
\midrule
1   & \underline{arp\_req(host\textsubscript{d})}     & \underline{host\textsubscript{s}} & switch\textsubscript{1}           & data \\
2   & pkt\_in(arp\_req(host\textsubscript{d}))        & switch\textsubscript{1}           & controller                        & control \\
3   & pkt\_out(arp\_req(host\textsubscript{d}),flood) & controller                        & switch\textsubscript{1}           & control \\
4   & arp\_req(host\textsubscript{d})                 & switch\textsubscript{1}           & switch\textsubscript{2}           & data \\
5   & pkt\_in(arp\_req(host\textsubscript{d}))        & switch\textsubscript{2}           & controller                        & control \\
6   & pkt\_out(arp\_req(host\textsubscript{d}),flood) & controller                        & switch\textsubscript{2}           & control \\
7   & arp\_req(host\textsubscript{d})                 & switch\textsubscript{2}           & host\textsubscript{2}             & data \\
8   & arp\_req(host\textsubscript{d})                 & switch\textsubscript{2}           & host\textsubscript{3}             & data \\
9   & arp\_req(host\textsubscript{d})                 & switch\textsubscript{2}           & switch\textsubscript{3}           & data \\
10  & pkt\_in(arp\_req(host\textsubscript{d}))        & switch\textsubscript{3}           & controller                        & control \\
11  & pkt\_out(arp\_req(host\textsubscript{d}),flood) & controller                        & switch\textsubscript{3}           & control \\
12  & \underline{arp\_req(host\textsubscript{d})}     & switch\textsubscript{3}           & \underline{host\textsubscript{d}} & data \\
13  & \underline{arp\_rep(host\textsubscript{d})}     & \underline{host\textsubscript{d}} & switch\textsubscript{3}           & data \\
14  & pkt\_in(arp\_rep(host\textsubscript{d}))        & switch\textsubscript{3}           & controller                        & control \\
15  & pkt\_out(arp\_rep(host\textsubscript{d}),port2) & controller                        & switch\textsubscript{3}           & control \\
16  & arp\_rep(host\textsubscript{d})                 & switch\textsubscript{3}           & switch\textsubscript{2}           & data \\
17  & pkt\_in(arp\_rep(host\textsubscript{d}))        & switch\textsubscript{2}           & controller                        & control \\
18  & pkt\_out(arp\_rep(host\textsubscript{d}),port3) & controller                        & switch\textsubscript{2}           & control \\
19  & arp\_rep(host\textsubscript{d})                 & switch\textsubscript{2}           & switch\textsubscript{1}           & data \\
20  & pkt\_in(arp\_rep(host\textsubscript{d}))        & switch\textsubscript{1}           & controller                        & control \\
21  & pkt\_out(arp\_rep(host\textsubscript{d}),port1) & controller                        & switch\textsubscript{1}           & control \\
22  & \underline{arp\_rep(host\textsubscript{d})}     & switch\textsubscript{1}           & \underline{host\textsubscript{s}} & data \\
\bottomrule
\end{tabularx}
\end{table*}

\textbf{Message sequences.}
In the control plane of an SDN, an SDN controller exchanges sequences of control messages with SDN switches to establish and manage communication among hosts, monitor network status, and enforce network policies.
In the data plane of an SDN, hosts exchange sequences of data messages through SDN switches to transmit and receive various types of data, such as audio and video streams.
For example, Table~\ref{tbl:message sequence} presents an example sequence of messages aimed at discovering host locations (i.e., MAC addresses~\cite{PetersonD2007:CompNet}) in the SDN network shown in Figure~\ref{fig:SDN example}.

Regarding the example sequence listed in Table~\ref{tbl:message sequence}, we consider an SDN setup in which the controller in Figure~\ref{fig:SDN example} is unaware of a path across switches that enables the transmission of data messages from host\textsubscript{s} to host\textsubscript{d}.
The address resolution protocol (ARP) is typically used to map an IP address of a host to its physical (MAC) address~\cite{Plummer1982:ARP, PetersonD2007:CompNet}.
The first ARP message $m_1$ generated by host\textsubscript{s} is an ARP request aimed at obtaining the MAC address of host\textsubscript{d}.
The ARP request reaches to switch\textsubscript{1} that is connected to host\textsubscript{s}.
The switch then sends the ARP request to the controller by encapsulating it through the packet-in control message $m_2$.
The controller is now aware of the information regarding the source of the ARP request, i.e, host\textsubscript{s}.
However, since the controller does not know the location of host\textsubscript{d}, it instructs switch\textsubscript{1} to flood the ARP request to the connected switches using the packet-out message $m_3$.
The ARP request is then flooded in the network (via $m_4$ to $m_{11}$) until it reaches the destination host\textsubscript{d} (via $m_{12}$).
The destination host\textsubscript{d} then sends the ARP reply $m_{13}$ to switch\textsubscript{3} in order to inform the source host\textsubscript{s} of its location (MAC).
Note that, at this stage, since the controller knows the location of host\textsubscript{s}, it directly instructs the three switches with the exact directions (i.e., port numbers) to forward the ARP reply (see $m_{14}$ to $m_{22}$).
After this procedure, the controller usually installs forwarding rules for both ARP and IP messages to the switches, resulting in different sequences of messages compared to the example sequence mentioned above.

\textbf{Failures.}
Like any software component, SDN systems are susceptible to failures that can affect their functionality.
These failures may result in service disruptions noticeable to users.
Numerous studies have examined these failures in the context of SDN testing~\cite{Lee2017:DELTA,Jero2017:BEADS, OllandoSB2023:FuzzSDN, LeeWKYPS20,ShuklaSSCZF20}.
Furthermore, the centralisation of SDN system logic within its controller makes it a critical point of failure.
A controller crash or loss of connection with the switches can disrupt the entire network.
This vulnerability underscores the necessity for thorough testing to ensure the system's robustness and reliability.
Such testing entails exploring the state space of the SDN system, including scenarios that are not easily reached.
Unfortunately, no work has yet investigated how to automate such state space exploration in SDN systems.

\textbf{Problem.}
SDN controllers are inherently stateful.
They manage complex states that encompass their internal states, the connected switches' states, and the overall network states.
When developing and operating SDN systems, engineers must address system failures triggered by unexpected sequences of control messages.
Specifically, they must ensure that the system behaves acceptably regardless of its current state.
In SDN systems, a stateful controller is susceptible to entering incorrect states, sending unexpected messages, or triggering system failures.
These failures may occur only when the controller, the connected switches, or the network reach specific states.
For instance, an erroneous control message may be handled correctly under nominal conditions, but if the same message is transmitted during a state of `recovery' of the system, a system failure may occur.
When such a failure occurs, engineers must determine the state of the controller at the time of the failure and identify the sequence of messages that led to that state.
Precisely identifying these conditions is crucial, as it enables engineers to diagnose the failure with a clear understanding of the conditions that caused it.
Additionally, engineers can utilise this information to generate extended sets of control message sequences for testing the system after implementing fixes.
Our work aims to efficiently and effectively test the controller of an SDN system by identifying control message sequences that cause failures, and then automatically derive an accurate model that characterises the sequences of messages leading to failures.

\section{Approach}
\label{sec:approach}

\subsection{Overview}
\label{subsec:overview}

\begin{figure}[t]
    \centering
    \includegraphics[width=0.75\textwidth]{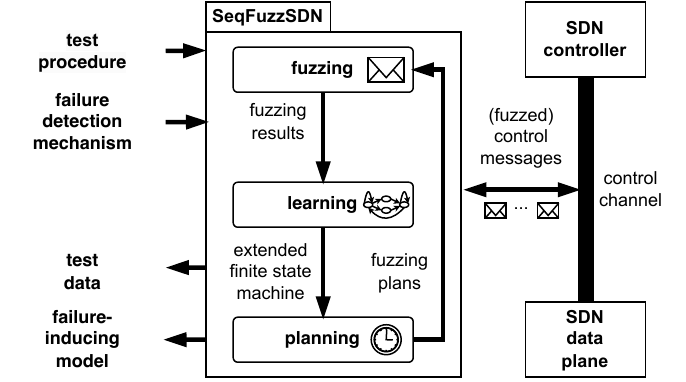}
    \caption{Approach overview. To test a stateful SDN controller, \FRAM fuzzes control message sequences guided by inferred extended finite state machines (EFSMs) that capture failure-inducing message sequences.}
    \label{fig:approach overview}
    \Description{}
\end{figure}

\Cref{fig:approach overview} shows an overview of \FRAM.
\FRAM takes as input a test procedure and a failure detection mechanism.
The test procedure specifies the steps required to (1)~initialise the controller under test, switches, and hosts in an SDN, (2)~execute a use scenario, e.g., pair-wise ping test~\cite{RFC1122:PING}, to test the controller, and (3)~properly tear down the SDN based on the given use scenario to test the controller again.
Note that depending on the given use scenario, sequences of control messages exchanged between the controller and switches can vary.
The failure detection mechanism, defined by engineers for the given test procedure, allows \FRAM to determine whether the controller fails.
For example, unexpected communication breakdowns and significant performance degradation can be considered as failures depending on the given test procedure.
Regarding the outputs of \FRAM, it produces a test data set and a failure-inducing model.
The former contains sequences of control messages that are fuzzed by \FRAM and lead to failures detected by the failure detection mechanism. 
The failure-inducing model characterises sequences of control messages leading to either successes or failures.
When the failure detection mechanism does not detect any failures, \FRAM considers the corresponding message sequences as successful.
In summary, \FRAM aims at generating a test data set that contains diverse failure-inducing sequences of control messages and a failure-inducing model that accurately characterises them.

\FRAM is an iterative fuzzing method consisting of three steps (see Figure~\ref{fig:approach overview}), as follows:
(1)~The fuzzing step involves sniffing and modifying control messages that pass through the control channel between the SDN controller and the SDN switches.
Hence, it does not require any changes to the SDN controller and switches.
(2)~The learning step takes as input the control message sequences and failure detection results obtained from the fuzzing step. 
The learning step then builds a model to characterise the message sequences.
Specifically, the learning step infers an extended finite state machine (EFSM)~\cite{AlagarP2011:EFSM} that captures the controller's behaviour in terms of state transitions representing control messages received or sent by the controller.
Unlike FSMs, EFSMs can capture state transitions associated with data variables, which are essential for modelling state changes caused by control messages.
Indeed, control messages typically involve control operations that depend on data (e.g., flow tables and packet statistics).
The inferred EFSM contains two types of final states representing success and failure, enabling \FRAM to classify and predict which sequences of state transitions (i.e., control messages) induce either success or failure.
(3)~The planning step takes as input the EFSM inferred by the learning step and generates fuzzing plans.
These fuzzing plans aim to guide the fuzzing step in efficiently exploring the possible space of control message sequences and discovering diverse failure-inducing sequences of control messages.
In the following subsections, we provide detailed descriptions of the three steps in \FRAM.

\subsection{Fuzzing}
\label{subsec:fuzzing}

The fuzzing step of \FRAM relies on the man-in-the-middle attack (MITM) technique~\cite{ContiDL2016}, which is widely used and studied in the network security domain.
This technique enables \FRAM to position itself between the controller under test and the SDN switches that are communicating with the controller.
Using MITM, \FRAM can intercept control messages and potentially fuzz them while ensuring that the controller and switches remain unaware of the presence of \FRAM.
Furthermore, employing this attack technique allows \FRAM to generate realistic potential threats (i.e., unexpected sequences of control messages) that the controller may face in practice.

When fuzzing control messages, \FRAM accounts for the syntax requirements (i.e., grammar) defined in an SDN protocol (e.g., OpenFlow) to ensure fuzzed control messages are syntactically valid.
An SDN controller typically rejects syntactically invalid messages at its message parsing layer~\cite{Jero2017:BEADS}.
Hence, producing valid control messages is desirable in practice to test the controller's behaviour beyond the parsing layer~\cite{Jero2017:BEADS, OllandoSB2023:FuzzSDN}.

\FRAM employs a mutation-based fuzzing strategy~\cite{fuzzingbook2024} in which fuzz (i.e., mutation) operators introduce small changes to sniffed control messages while adhering to the syntax requirements of the messages.
Below, we first describe five fuzz operators employed in \FRAM that can modify control messages and their sequences.
We then describe in detail how \FRAM uses the fuzz operators.

\subsubsection{Fuzz Operators.}
\label{subsubsec:fuzz operators}

\begin{figure}[t]
    \centering
    \includegraphics[width=\columnwidth]{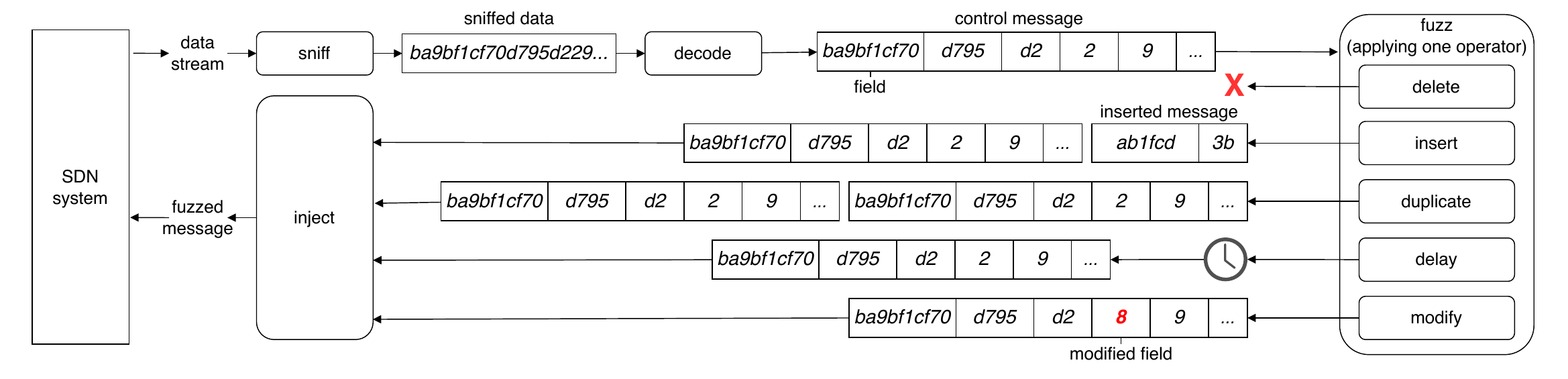}
    \caption{A data flow example of fuzzing by applying either the deletion, insertion, duplication, delay, or modification operator.}
    \label{fig:data flow example}
    \Description{}
\end{figure}

As shown in Figure~\ref{fig:data flow example}, when \FRAM sniffs a control message, it can apply one of the following fuzz operators: deletion, insertion, duplication, delay, and modification.
These operators are based on those used in \BEADS~\cite{Jero2017:BEADS}, with modifications tailored for the learning-guided fuzzing of \FRAM.
We describe further details of the fuzz operators below.

\noindent\textbf{Deletion.}
The deletion operator drops an intercepted message.
For example, when \FRAM intercepts a packet-in message from the control channel, it can omit retransmitting the message to the channel, thereby deleting the packet-in message from the control channel.

\noindent\textbf{Insertion.}
The insertion operator inserts a new control message into the control channel.
For example, \FRAM can insert a new packet-in message to the control channel while it sniffs messages passing through the channel.
Note that, such a new message is either predefined by engineers or randomly generated, as configured in \FRAM. 

\noindent\textbf{Duplication.}
The duplication operator duplicates a sniffed message.
For example, when \FRAM intercepts a packet-in message, it can copy the same message and resend both the original and copied messages to the control channel.
Hence, the channel carries the duplicated packet-in messages.

\noindent\textbf{Delay.}
The delay operator holds a control message for a certain amount of time.
For example, \FRAM can hold an intercepted packet-in message for 200ms and resend it after the delay time.
When \FRAM holds a synchronous message (e.g., barrier-request), the sender will also wait for a response from the receiver.
However, if \FRAM delays an asynchronous message (e.g., packet-in), the sender continues its processing without waiting for the receiver to respond.
Note that the delay time can be configured in \FRAM.

\noindent\textbf{Modification.}
The modification operator modifies the content (i.e., fields) of an intercepted control message.
For example, when \FRAM intercepts a packet-in message, it can change the version field of the message and inject the fuzzed message into the control channel.

\begin{algorithm}[t]
\caption{Modification: Syntax-aware random}
\label{alg:modification random}
\begin{algorithmic}[1]
\Input
\Statex $\mathit{msg}$: control message to be fuzzed
\Statex $\mathit{pf}$: probability of fuzzing a field
\Output
\Statex $\mathit{msg^\prime}$: control message after fuzzing
\Statex

\State $F \gets \Call{get\_fields}{msg}$
\State $\mathit{msg}^\prime \gets \mathit{msg}$
\ForAll{$f \in F$}
    \If{\Call{rand}{$0, 1$} $\le \mathit{pf}$}
        \State $\mathit{msg}^\prime \gets \Call{replace}{\mathit{msg^\prime}, f, \textproc{rand\_valid}(f)}$
    \EndIf
\EndFor
\State \Return $\mathit{msg^\prime}$
\end{algorithmic}
\end{algorithm}

We note that the modification operator behaves differently for the initial fuzzing phase and the subsequent learning-guided fuzzing phases.
Algorithm~\ref{alg:modification random} shows how the modification operator functions during the initial fuzzing phase when there is no guidance available for fuzzing.
Given an intercepted message $\mathit{msg}$, the modification operator parses the content of $\mathit{msg}$ in terms of its fields (line~1).
For each field $f$ of $\mathit{msg}$ and a given probability $\mathit{pf}$ of fuzzing a field, the operator replaces its value with a random value within its syntactically valid value range (lines~2-7).
For example, if $\mathit{msg}$ contains ten fields and $\mathit{pf} = 0.2$, after applying the operator, the expected number of modified fields in $\mathit{msg}$ is two.
The operator then returns the fuzzed message $\mathit{msg^\prime}$ to transmit it through the control channel.

In the subsequent iterations of \FRAM, the modification operator leverages planning outputs obtained from the learning and planning steps (see Figure~\ref{fig:approach overview}).
For readability, we describe the modification operator guided by learning in Section~\ref{subsec:guided fuzzing}, after introducing the learning and planning steps.

\subsubsection{Initial Fuzzing.}
\label{subsubsec:initial fuzzing}

\begin{algorithm}[t]
\caption{Initial fuzzing}
\label{alg:initial fuzzing}
\begin{algorithmic}[1]
\Input
\Statex $\mathit{pm}$: probability of fuzzing a message
\Output
\Statex $\mathit{seq^\prime}$: sequence of messages after fuzzing
\Statex

\State $\mathit{seq^\prime} \gets \langle \rangle$
\Repeat
    \State $\mathit{msg} \gets \Call{receive}$
    \State \textcolor{gray}{// no fuzzing}
    \If{$\Call{rand}{0, 1} > \mathit{pm}$}
        \State $\Call{send}{\mathit{msg}}$
        \State $\mathit{seq^\prime} \gets \Call{append}{\mathit{seq^\prime}, \mathit{msg}}$
        \State \Continue
    \EndIf
    \State \textcolor{gray}{// fuzzing}
    \State $\mathit{op} \gets \Call{rand\_select\_fuzz\_operator}$
    \If{$\mathit{op}$ is a deletion operator}
        \State \textcolor{gray}{// do nothing}
    \ElsIf{$\mathit{op}$ is an insertion operator}
        \State $\mathit{msg^\prime} \gets \Call{get\_message}{\mathit{op}}$
        \State $\Call{send}{\mathit{msg}, \mathit{msg^\prime}}$
        \State $\mathit{seq^\prime} \gets \Call{append}{\mathit{seq^\prime}, \mathit{msg}, \mathit{msg^\prime}}$
    \ElsIf{$\mathit{op}$ is a duplication operator}
        \State $\Call{send}{\mathit{msg}, \mathit{msg}}$
        \State $\mathit{seq^\prime} \gets \Call{append}{\mathit{seq^\prime}, \mathit{msg}, \mathit{msg}}$
    \ElsIf{$\mathit{op}$ is a delay operator}
        \State $t \gets \Call{get\_delay}{\mathit{op}}$
        \State $\Call{delay\_send}{\mathit{msg}, t}$
        \State $\mathit{seq^\prime} \gets \Call{delay\_append}{\mathit{seq^\prime}, \mathit{msg}, t}$
    \ElsIf{$\mathit{op}$ is a modification operator}
        \State $\mathit{msg^\prime} \gets \Call{modify}{\mathit{msg}, \mathit{op}}$
        \State $\Call{send}{\mathit{msg^\prime}}$
        \State $\mathit{seq^\prime} \gets \Call{append}{\mathit{seq^\prime}, \mathit{msg^\prime}}$
    \EndIf
\Until{the test procedure has finished}

\State \Return $\mathit{seq^\prime}$
\end{algorithmic}
\end{algorithm}

During the initial fuzzing phase of \FRAM, since no failure-inducing model has been inferred, \FRAM applies the fuzz operators randomly, as described in Algorithm~\ref{alg:initial fuzzing}.
The algorithm takes as input a probability $\mathit{pm}$ of fuzzing a message.
It then returns a sequence $\mathit{seq^\prime}$ of messages after fuzzing.
While executing a given test procedure (see the repeat block on lines 2-30 in Algorithm~\ref{alg:initial fuzzing}), \FRAM intercepts each of the control messages passing through the control channel between the SDN controller and switches (line~3).
For each control message $\mathit{msg}$ and the given probability $\mathit{pm}$, \FRAM decides whether it fuzzes $\mathit{msg}$ or not (line~5).
When \FRAM does not fuzz $\mathit{msg}$, it resends $\mathit{msg}$ to the control channel and appends $\mathit{msg}$ to $\mathit{seq^\prime}$ to record a processed message sequence (lines~6-7).
For fuzzing the message $\mathit{msg}$, \FRAM randomly selects one of the five fuzz operators (line~11).
\FRAM then applies the selected operator to $\mathit{msg}$ and updates $\mathit{seq^\prime}$ accordingly (lines~12-29).

\subsubsection{Data Collection.}
\label{subsubsec:data collection}

To generate failure-inducing models, \FRAM relies on an inference technique that takes as input event traces and produces an extended finite state machine (EFSM), such as \textsc{Mint}~\cite{Walkinshaw2016:MINT}.
This EFSM captures the event traces as state transitions with guard conditions.
In our context, an event trace corresponds to a message sequence $\mathit{seq^\prime}$ obtained from the fuzzing step.
Each event $e$ in the trace is associated with the corresponding message $\mathit{msg}$ listed in $\mathit{seq^\prime}$. 
Specifically, an event $e$ is a tuple $(l, m, v)$, where $l$ denotes the type of $\mathit{msg}$, $m$ denotes the fuzz operator applied to $\mathit{msg}$, $v$ denotes the field values of $\mathit{msg}$.
Note that $m$ can be null if $\mathit{msg}$ is not fuzzed in the given message sequence $\mathit{seq^\prime}$.
For example, consider a control message sequence in which a hello control message~\cite{OpenFlowSpec}, used to discover and establish a connection between the controller and switches, was delayed by 200ms using the delay operator.
\FRAM encodes this hello message into an event $e$ as follows: $(\mathtt{hello},~\mathtt{delay}:200,~\mathtt{<0x5,0x0,0x10,0xA34BF>})$, where the field values of the hello message are version $=$ $\mathtt{0x5}$, type $=$ $\mathtt{0x0}$, length $=$ $\mathtt{0x10}$, and xid $=$ $\mathtt{0xA34BF}$.
The last event in the trace indicates either success or failure, determined by the failure detection mechanism for the given sequence $\mathit{seq^\prime}$ of messages.
In addition, the event $e$ is associated with both the sender and receiver of $\mathit{msg}$, enabling \FRAM to track this information.

We note that, at each iteration $i$ of \FRAM, the fuzzing step executes the input test procedure (see Figure~\ref{fig:approach overview}) $n$ times, determined by a time budget.
Hence, for each iteration $i$, the fuzzing step generates a dataset $D_i$ that contains $n$ event traces, i.e., $|D_i|$ $=$ $n$.

\subsection{Learning}
\label{subsec:learning}

At each iteration $i$ of \FRAM, the learning step takes as input a dataset $D$ of event traces obtained from the fuzzing step through the $1$st to $i$th iterations, i.e., $D$ $=$ $D_1$ $\cup$ $\ldots$ $\cup$ $D_i$.
The learning step then outputs an EFSM inferred based on $D$.
The inferred EFSM $M$ is then used to guide the fuzzing process, which entails exploiting state transitions in $M$, exploring less-visited states in $M$, and discovering new states not captured in $M$.
Furthermore, \FRAM provides engineers with an accurate EFSM, achieved through iterative refinement of $M$.
This EFSM serves as a failure-inducing model that characterises the generated failure-inducing message sequences (i.e., event traces), enabling engineers to gain a more comprehensive understanding of failure-inducing sequences rather than individually inspecting each of them.

We note that, to infer an EFSM, \FRAM relies on \textsc{Mint}~\cite{Walkinshaw2016:MINT}, a state-of-the-art model inference tool that takes as input a dataset containing event traces and produces an EFSM.
We opted to use \textsc{Mint} since it is one of the few tools available online and has been applied in many software engineering studies~\cite{ShinBB2022:PRINS, EmamM2018:ReHMM}.
In addition, the implementation of \textsc{Mint} is the most reliable among the tools available online, enabling us to focus on developing the main contributions of \FRAM.

\subsubsection{States and transitions.}
\label{subsubsec:states and transitions}

An SDN controller takes as input control messages and produces control messages in response, which are observable via MITM techniques~\cite{ContiDL2016}.
In an EFSM inferred by \FRAM, which captures sequences of observed control messages, a state is a placeholder for the transitions between different sequences of these messages, rather than representing a specific internal condition of the controller, which is not visible to \FRAM.
In this state, the controller is capable of processing a particular control message and generating a corresponding response message.
A transition is defined as a tuple $(s, l, c, m, d)$, where $s$ denotes a source state, $l$ denotes the type of a control message, $c$ denotes a guard condition on the fields of the message, $m$ denotes a fuzz operator applied to the message, and $d$ denotes a destination state.
In an EFSM produced by \FRAM, using the dataset $D$ containing event traces, each transition $(s, l, c, m, d)$ corresponds to an event $(l, m, v)$ in $D$ (see the event definition in Section~\ref{subsubsec:data collection}). 

\begin{figure}[t]
    \centering
    \includegraphics[width=0.9\linewidth]{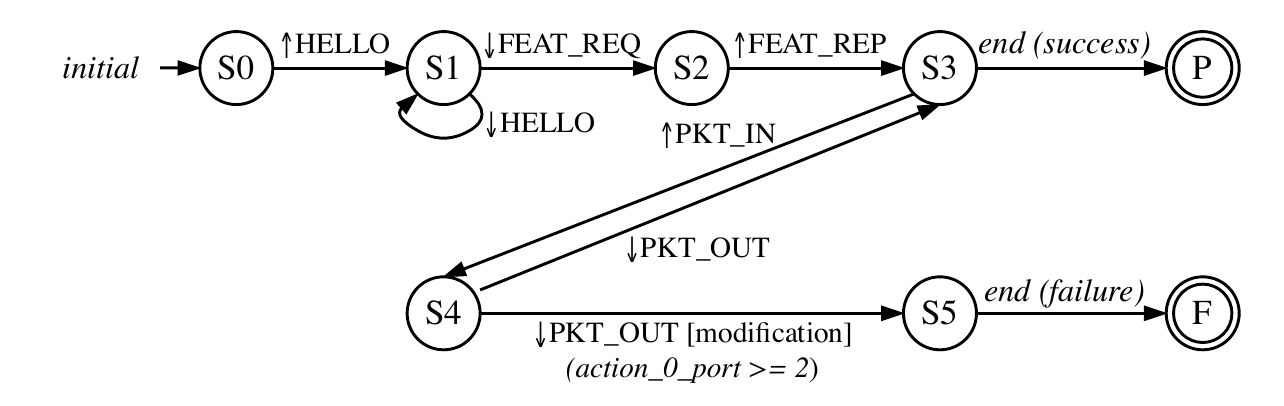}
    \caption{A simplified EFSM example produced by \FRAM. The $\uparrow$ and $\downarrow$ arrows indicate that the corresponding control messages are received and sent, respectively, by the controller under test.}
    \label{fig:simplified efsm}
    \Description{}
\end{figure}

For example, Figure~\ref{fig:simplified efsm} shows an EFSM produced by \FRAM, simplified for clarity.
The EFSM contains eight states in total.
Among these states, state \texttt{S0} represents the initial state of the EFSM, state \texttt{P} represents the success state, and state \texttt{F} represents the failure state.
Additionally, the EFSM includes ten transitions.
For instance, in the transition from state \texttt{S4} to state \texttt{S5}, \texttt{S4} serves as the source state ($s$), \texttt{PACKET\_OUT} as the label ($l$), $\mathtt{action\_0\_port} \geq 2$ as the guard condition ($c$), \texttt{modification} as the mutation operator ($m$), and \texttt{S5} as the destination state ($d$).
Note that the arrow $\uparrow$ (resp. $\downarrow$) annotated before each label (e.g., $\uparrow$\texttt{HELLO} and $\downarrow$\texttt{HELLO}) indicates that a message is received by the controller (resp. sent by the controller).

\subsubsection{Guard condition inference.}
\label{subsubsec:guard condition inference}

\FRAM aims at efficiently producing an accurate EFSM that correctly captures the event traces $D$.
Since the overall accuracy of an EFSM highly depends on the accuracy of transitions' guard conditions, in this section, we first explain how \FRAM uses \textsc{Mint} to infer guard conditions from $D$.
For further details, such as merging states and removing non-determinism, we refer readers to the paper introducing \textsc{Mint}~\cite{Walkinshaw2016:MINT}.
We then introduce how \FRAM efficiently infers an EFSM from $D$ in Section~\ref{subsubsec:Sampling}.

\textsc{Mint} employs a supervised machine learning algorithm~\cite{WittenFH2016} that requires labelled datasets to infer guard conditions of state transitions.
To create labelled datasets, \FRAM groups events in the event traces $D$ into event groups $E$ based on the event type defined by $(l, m)$, where the control message type $l$ and the fuzz operator $m$ are elements in an event $e$ $=$ $(l, m, v)$.
This ensures that each group contains events with the same event type.
For each event $e$ in an event group, \FRAM then labels $e$ with the type of the next event following $e$ in the corresponding event trace of $e$ in the event traces $D$, as required by \textsc{Mint}.
We note that one of the key reasons why \textsc{Mint} labels each event with the next event type is to enable machine learning classifiers to learn the guards that govern transitions between states, specifically identifying the conditions under which a state that allows a given event can transition to another state that allows an event of the next type.
More precisely, given an event trace $e_1$, $\ldots$, $e_i$, $e_{i+1}$, $\ldots$, $e_n$, the event $e_i$ in an event group is labelled with the type $(l_{i+1},m_{i+1})$ of $e_{i+1}$.
For each event group $E$, \FRAM creates a dataset that contains pairs of the field values of an event $e \in E$ and the assigned label of $e$.  

\begin{table*}[t]
\caption{An example illustrating the creation of datasets based on event traces:
(a)~Two event traces (i.e., Trace 1 and Trace 2). (b)~Six datasets created based on Trace 1 and Trace 2.}
\label{tab:traces to datasets}
\begin{subtable}[c]{\textwidth}
\centering
\small
\subcaption{event traces}
\begin{tabularx}{\textwidth}{cYYYYY}
\toprule
& event (e) & \multicolumn{1}{c}{label (l)} & \multicolumn{1}{c}{mutation (m)} & \multicolumn{1}{c}{value (v)} \\
\midrule
\multirow{7}{*}{Trace 1}
  & $e_{10}$ & HELLO & null & $v(e_{10})$ \\
  & $e_{11}$ & HELLO & null & $v(e_{11})$ \\
  & $e_{12}$ & FEAT\_REQ & null & $v(e_{12})$ \\
  & $e_{13}$ & FEAT\_REP & null & $v(e_{13})$ \\
  & $e_{14}$ & PKT\_IN & null & $v(e_{14})$ \\
  & $e_{15}$ & PKT\_OUT & null & $v(e_{15})$ \\
  & $e_{16}$ & end(success) & null & null \\
\arrayrulecolor{lightgray}\midrule\arrayrulecolor{black}
\multirow{7}{*}{Trace 2}
  & $e_{21}$ & HELLO & null & $v(e_{21})$ \\
  & $e_{22}$ & HELLO & null & $v(e_{22})$ \\
  & $e_{23}$ & FEAT\_REQ & null & $v(e_{23})$ \\
  & $e_{24}$ & FEAT\_REP & null & $v(e_{24})$ \\
  & $e_{25}$ & PKT\_IN & null & $v(e_{25})$ \\
  & $e_{26}$ & PKT\_OUT & modification & $v(e_{26})$ \\
  & $e_{27}$ & end(failure) & null & null \\
\bottomrule
\end{tabularx}
\end{subtable}
\hfill \vspace{1em}
\begin{subtable}[c]{\textwidth}
\small
\centering
\subcaption{datasets}
\label{subtab:dataset}
\begin{tabularx}{\textwidth}{l@{\hspace{-1em}}Y@{\hspace{-1em}}Y@{\hspace{-1em}}Y}
\toprule
\multicolumn{1}{c}{event type $(l, m)$} & \multicolumn{1}{c}{event (e)} & \multicolumn{1}{c}{value (v)} & \multicolumn{1}{c}{next event type $(l^\prime,m^\prime)$} \\
\midrule
\multirow{4}{*}{(HELLO, null)}
  & $e_{10}$ & $v(e_{10})$ & (HELLO, null) \\
  & $e_{11}$ & $v(e_{11})$ & (FEAT\_REQ, null) \\
  & $e_{21}$ & $v(e_{21})$ & (HELLO, null) \\
  & $e_{22}$ & $v(e_{22})$ & (FEAT\_REQ, null) \\
  \arrayrulecolor{lightgray}\midrule\arrayrulecolor{black}
\multirow{2}{*}{(FEAT\_REQ, null)}
  & $e_{12}$ & $v(e_{12})$ & (FEAT\_REP, null) \\
  & $e_{23}$ & $v(e_{23})$ & (FEAT\_REP, null) \\	
  \arrayrulecolor{lightgray}\midrule\arrayrulecolor{black}

\multirow{2}{*}{(FEAT\_REP, null)}
  & $e_{13}$ & $v(e_{13})$ & (PKT\_IN, null) \\
 & $e_{24}$ & $v(e_{24})$ & (PKT\_IN, null) \\
 \arrayrulecolor{lightgray}\midrule\arrayrulecolor{black}

\multirow{2}{*}{(PKT\_IN, null)}
  & $e_{14}$ & $v(e_{14})$ & (PKT\_OUT, null) \\
  & $e_{25}$ & $v(e_{25})$ & (PKT\_OUT, modification) \\
  \arrayrulecolor{lightgray}\midrule\arrayrulecolor{black}

(PKT\_OUT, null)
  & $e_{15}$ & $v(e_{15})$ & (end(success), null) \\
  \arrayrulecolor{lightgray}\midrule\arrayrulecolor{black}

(PKT\_OUT, modification)
  & $e_{26}$ & $v(e_{26})$ & (end(failure), null) \\
\bottomrule
\end{tabularx}
\end{subtable}
\end{table*}

For example, Table~\ref{tab:traces to datasets} shows the creation of six datasets (see Table~\ref{tab:traces to datasets}~(b)) from two event traces (see Table~\ref{tab:traces to datasets}~(a)).
The datasets could be used to infer the EFSM presented in Figure~\ref{fig:simplified efsm}.
As shown in Table~\ref{tab:traces to datasets}~(a), Trace 1 and Trace 2 contain six distinct event types ($l$, $m$): ($HELLO$, $\mathtt{null}$), ($FEAT\_REQ$, $\mathtt{null}$), ($FEAT\_REP$, $\mathtt{null}$), ($PKT\_IN$, $\mathtt{null}$), ($PKT\_OUT$, $\mathtt{null}$), and ($PKT\_OUT$, $\mathtt{modification}$).
For each event type, \FRAM creates its corresponding dataset as presented in Table~\ref{tab:traces to datasets}~(b).
Note that the third and last columns in the table indicate the content of a labelled dataset, including field values of a control message and associated labels (i.e., the next event type).

Regarding supervised machine learning, \FRAM relies on RIPPER (Repeated Incremental Pruning to Produce Error Reduction)~\cite{Cohen1995:FERI}, an interpretable rule-based classification algorithm. 
RIPPER has shown successful applications in many software engineering problems involving classification and condition inference~\cite{HaqSNB21, Brindescu0LS20, LuoNFGP2017:FOREPOST}.
In particular, we select RIPPER because it generates pruned decision rules (i.e., if-conditions) that are more concise and, as a result, more interpretable than commonly used tree-based classification algorithms, such as C4.5~\cite{Quinlan1993:C45}, which are susceptible to the replicated subtree problem~\cite{WittenFH2016}.

\subsubsection{Sampling Event Traces}
\label{subsubsec:Sampling}

Due to the computational complexity of the model inference problem, existing model inference techniques (including \textsc{Mint}) face scalability issues~\cite{ShinBB2022:PRINS}.
Among the works addressing the scalability problem, \citet{ShinBB2022:PRINS} recently introduced \textsc{PRINS}, which is the most relevant to \textsc{MINT}, the EFSM inference tool we selected for \FRAM.
\textsc{PRINS} aims to improve the scalability of EFSM inference for component-based systems.
It employs a divide-and-conquer approach, first deriving individual models for each system component based on their respective logs.
These component models are then systematically merged, incorporating the event flow across components as recorded in the logs.
However, \textsc{PRINS} is not suitable for \FRAM, as it requires prior knowledge of which system components generate logs.

In the context of \FRAM, the number of event traces can continuously grow as \FRAM iterates through the fuzzing, learning, and planning steps multiple times and fuzzes message sequences corresponding to those event traces.
Hence, when the number of event traces and their events become large (e.g., 5000 event traces, containing 15000 events), \textsc{Mint} either crashes due to running out of memory or takes a prohibitively long time to complete its execution.
To address the scalability problem in our context, when learning an EFSM at each iteration $i$, \FRAM uses a subset $D^s$ of event traces instead of all event traces $D$ generated up to the current iteration.

Let $M$ be an EFSM inferred at the $i{-}1$th iteration of \FRAM.
At iteration $i$, to sample event traces from the event traces $D_i$ obtained at $i$ and those used in learning $M$, \FRAM first separate $D_i$ into \emph{accepted} and \emph{rejected} traces. 
Given an EFSM $M$, accepted traces are traces that are already explained by $M$, i.e., traces that follow paths (i.e., transition sequences) in $M$.
Note that when guard evaluations are needed while \FRAM walks over $M$ with traces, it uses Z3~\cite{DeMoura2008:Z3}, a well-known and widely applied SMT solver.
In contrast, rejected traces refer to traces that do not follow any path in $M$.
Hence, to create a set of event traces for learning a new EFSM $M^\prime$ at iteration $i$, \FRAM includes the rejected traces in the set in order to ensure that they are explained by $M^\prime$.
However, \FRAM does not include the accepted traces in the learning process because they do not contribute to refining $M$ into $M^\prime$.

\FRAM then further separates the rejected event traces obtained from iteration $i$ of \FRAM into success event traces and failure event traces.
Drawing inspiration from the observation that balanced datasets often yield higher accuracy in ML~\cite{WittenFH2016,BettaiebSSBNG19,Bettaieb20}, \FRAM manages two distinct sets of event traces: one leading to success and the other to failure.
These sets have the same maximum number of event traces and are used together to learn an EFSM.

\begin{algorithm}[t]
	\caption{Sampling event traces. Note that the sets of event traces used in this algorithm contain only success or failure event traces.}
	\label{alg:sampling event traces}
	\begin{algorithmic}[1]
		\Input
		\Statex $M$: EFSM
		\Statex $\mathcal{D}^M$: set of success (resp. failure) event traces used to generate $M$
		\Statex $\mathcal{D}_i$: set of success (resp. failure) event traces obtained from the $i$th iteration of \FRAM
		\Statex $n_{\mathcal{D}}$: maximum size of an output set of event traces
		
		\Output
		\Statex $\mathcal{D}$: set of success (resp. failure) event traces for learning a new EFSM
		\Statex

		\Comment{Case: include all traces}		
		\If{$\card{\mathcal{D}^M \cup \mathcal{D}_i} \leq n_{\mathcal{D}}$}
		\State $\mathcal{D}$ $\gets$ $\mathcal{D}^M$ $\cup$ $\mathcal{D}_i$
		\State \Return $\mathcal{D}$
		\EndIf		
		\State
		\Comment{Case: replace traces}
		\State $n_r$ $\gets$ $\card{\mathcal{D}^M \cup \mathcal{D}_i} - n_{\mathcal{D}}$ {\color{gray}// number of traces to replace}
		\For{$n_r$ \textbf{times}}
		\State $\mathbb{G}$ $\gets$ \Call{group\_by\_path}{$\mathcal{D}^M$, $M$} {\color{gray}// $\mathbb{G}$: set of trace groups}
		\State $G$ $\gets$ \Call{select\_max\_group}{$\mathbb{G}$}
		\State 	$t$ $\gets$ \Call{rand\_select\_trace}{$G$}
		\State $\mathcal{D}^M$ $\gets$ $\mathcal{D}^M$ $\setminus$ $\{t\}$				
		\EndFor
		\State $\mathcal{D}$ $\gets$ $\mathcal{D}^M$ $\cup$ $\mathcal{D}_i$	 {\color{gray}// $\card{\mathcal{D}} = n_{\mathcal{D}}$}			
		\State \Return $\mathcal{D}$
	\end{algorithmic}
\end{algorithm}

Algorithm~\ref{alg:sampling event traces} presents our heuristic for sampling event traces.
\FRAM applies the algorithm separately to both success rejected event traces and failure rejected event traces.
The algorithm takes as input an EFSM $M$ inferred at iteration $i{-}1$, a set $\mathcal{D}^M$ of success (resp. failure) event traces used to learn $M$, a set $\mathcal{D}_i$ of success (resp. failure) rejected event traces obtained from iteration $i$, and the maximum size $n_{\mathcal{D}}$ of an output set $\mathcal{D}$.
The algorithm then outputs a set $\mathcal{D}$ of success (resp. failure) event traces for learning a new EFSM.
As shown on lines 1--5 of the algorithm, when the size of $\mathcal{D}^M \cup \mathcal{D}_i$ does not exceed the maximum size  $n_{\mathcal{D}}$, the algorithm returns $\mathcal{D}^M \cup \mathcal{D}_i$.
Otherwise, on line 8, the algorithm computes the number $n_r$ of event traces to remove from $\mathcal{D}^M$ to ensure that the output set $D$ contains $n_\mathcal{D}$ event traces (see line 15).
On lines 9-14, the algorithm removes $n_r$ event traces from $\mathcal{D}^M$ as follows:
It first partitions $\mathcal{D}^M$ into groups, each containing event traces that follow the same path in $M$.
It then selects a group $G$ that contains the largest number of event traces compared to the other groups.
On lines 12-13, it randomly selects an event trace $t$ and removes it from $\mathcal{D}^M$.
On lines 15-16, the algorithm returns $\mathcal{D}^M \cup \mathcal{D}_i$, where $\card{\mathcal{D}} = n_{\mathcal{D}}$.
Note that the selection mechanism on lines 10-11 aims at minimising information loss in $\mathcal{D}$ with regard to learning an EFSM.
Since the selection mechanism (lines 10-11) selects an event trace from group $G$ containing the largest number of event traces and removes the selected trace from $\mathcal{D}^M$ (lines 12-13), the remaining traces in $G$ will still contribute to creating a new EFSM that contains the same path (i.e., no information loss after the removal) and accepts the remaining traces.

\subsection{Planning}
\label{subsec:planning}

The planning step of \FRAM takes as input an EFSM and outputs fuzzing plans to guide the subsequent fuzzing iteration.
The fuzzing plans are defined as sequences of state transitions, i.e., paths in an EFSM, that guide the fuzzing step at the subsequent iteration. 
\FRAM produces the fuzzing plans, aiming at (O1)~exploring less-visited or new states of the controller under test, (O2)~improving the accuracy of a failure-inducing model (i.e., EFSM) and (O3)~increasing the diversity of message sequences (i.e., event traces) exercised for testing the controller.
Hence, \FRAM employs a multi-objective search algorithm~\cite{Deb2001} to address the planning problem.
Below, we describe the multi-objective search-based planning approach in \FRAM by defining the solution representation, the fitness functions, and the search algorithm.

\subsubsection{Representation}
\label{subsubsec:representation}

Given an EFSM $M$ obtained from the learning step, a candidate solution is a set $C$ of sequences of state transitions (i.e., paths) in $M$ where each transition sequence starts from the initial state $s_1$ of $M$ and ends at a state $s_o$ selected during search, representing a valid traversal of $M$.
Depending on a fuzzing probability, each transition sequence in $C$ can be associated with a fuzz operator $m_o$---deletion, insertion, duplication, delay, or modification described in Section~\ref{subsubsec:fuzz operators}---to be applied when the controller's state is $s_o$ in the subsequent iteration of \FRAM.

\subsubsection{Fitness Functions}
\label{subsubsec:fitness functions}

\FRAM aims at searching for candidate solutions with regard to the three objectives: (O1)~coverage, (O2)~accuracy, and (O3)~diversity, described earlier.
To quantify how a candidate solution fits these three objectives, below we define three fitness functions.

\paragraph{Coverage.}
\FRAM relies on an EFSM $M$ that models the state changes of the controller under test.
To test various behaviours of the controller, \FRAM aims at finding a candidate solution that ensures a similar (ideally equal) number of visits to each state in $M$.
Hence, each state in $M$ can be explored in different ways regarding how is reached and what happens after traversing it.
Given an EFSM $M$ at iteration $i$ of \FRAM and a set $D$ of event traces obtained from the first to the $i$th iterations, \FRAM counts the number of visits for each state in $M$ by traversing $M$ using each event trace in $D$.

To quantify the extent to which a candidate solution $C$ satisfies the coverage objective regarding the state-visit numbers, \FRAM leverages Shannon's Entropy~\cite{Shannon1948:Entropy}.
In general, entropy characterises the average level of uncertainty inherent to the stochastic variable's possible outcomes. 
In our context, the entropy defines the level of uncertainty associated with visits to a state in an EFSM $M$.
Intuitively, the higher the entropy, the more evenly the states in $M$ are visited.

Let $S$ be a set of all states in an EFSM $M$ obtained at iteration $i$ and $D$ be a set of event traces obtained from the first to the $i$th iterations.
For each state $s \in S$, we denote by $\mathit{nv}(s,C)$ the sum of the following: (1)~the number of visits to $s$ by the event traces in $D$, and (2)~the number of visits to $s$ expected by a candidate solution $C$.
We denote by $\mathit{nv}(S,C)$ the total number of state visits for $S$ and define $\mathit{nv}(S)$ $=$ $\sum_{s \in S}\mathit{nv}(s,C)$.
Based on Shannon's entropy equation, we formulate the following fitness function $\mathit{fitcov}(S,C)$ for the coverage objective as below.
\FRAM aims at maximising the fitness $\mathit{fitcov}(S,C)$.
\begin{equation*}
    \mathit{fitcov}(S,C) = -\sum\limits_{s \in S}\frac{\mathit{nv}(s,C)}{\mathit{nv}(S,C)}\log_2 \frac{\mathit{nv}(s,C)}{\mathit{nv}(S,C)}
\end{equation*}

We note that, in practice, an EFSM inference technique is not always able to infer an EFSM $M$ that allows the traversal of all event traces in $D$~\cite{Walkinshaw2016:MINT}.
Hence, \FRAM computes $\mathit{nv}(s,C)$ using those event traces in $D$ that are traceable by $M$ and a candidate solution $C$.
To improve the accuracy of an inferred EFSM over iterations of \FRAM, it accounts for an additional fitness function described below. 

\paragraph{Accuracy.}
\FRAM builds an EFSM $M$ using \textsc{Mint}, which relies on supervised machine learning.
Recall from Section~\ref{subsec:learning} that \textsc{Mint} converts the event traces $D$ into labelled training datasets (i.e., event groups) for building supervised classifiers.
Hence, building accurate classifiers is beneficial to improve the overall accuracy of an EFSM $M$.

Note that the imbalance problem~\cite{WittenFH2016} is one of the main reasons that usually cause the low performance of supervised classification algorithms.
In a labelled dataset, when the number of data instances of a class is significantly different from that of the other classes, classification algorithms tend to favour predicting the majority class, which is often not desirable in practice~\cite{WittenFH2016}.
Hence, \FRAM aims to address imbalance by planning to generate control message sequences that alleviate the problem.

To quantitatively assess the imbalance problem of each event group $E$ (i.e., labelled training dataset) converted from the event traces $D$, \FRAM uses the multi-class imbalance metric~\cite{LorenaGSKH2020}.
Given an event group $E$ obtained from the event traces $D$, we denote by $\mathit{nc}(E)$ the number of classes in $E$, $\mathit{ni}(E)$ the number of data instances in $E$, and $\mathit{ni}(c)$ the number of data instances labelled with the class $c$.
According to the multi-class imbalance metric, the imbalance ratio $\mathit{ir}(E)$ of $E$ is computed as follows:
\begin{equation*}
    \mathit{ir}(E) = \frac{\mathit{nc}(E) - 1}{\mathit{nc}(E)}\sum\limits_{c~\mathtt{in}~E}\frac{\mathit{ni}(c)}{\mathit{ni}(E) - \mathit{ni}(c)}
\end{equation*}

For example, consider an event group $E$ that consists of three classes ($\mathit{nc}(E) = 3$)---namely $c_1, c_2, c_3$---along with a total of 1200 data instances ($ni(E) = 1200$).
In the case where the class distribution is balanced (i.e., $\mathit{ni}(c_1) = \mathit{ni}(c_2) = \mathit{ni}(c_3) = 400$), the imbalance ratio is $\mathit{ir}(E) = 1$.
However, in a situation where the class distribution is imbalanced, such as $\mathit{ni}(c_1) = 5$, $\mathit{ni}(c_2) = 200$, $\mathit{ni}(c_3) = 995$, the imbalance ratio increases to $\mathit{ir}(E) \approx 3.37$.

To estimate the degree to which a candidate solution $C$ (i.e., fuzzing plan) impacts the imbalance problem, \FRAM augments each event group $E$ obtained from the event traces $D$ using $C$.
Recall from Section~\ref{subsec:learning} that each event group $E$ contains events that have the same event type. 
The class assigned to an event $e$ in $E$ is determined by the event following $e$ in the corresponding event trace (i.e., message sequence) containing $e$.
Hence, we can estimate how many new events will be added to each event group when \FRAM generates message sequences guided by a candidate solution $C$.
Precisely, given a sequence $(s_1,l_1,m_1,c_1,d_1)$, $\ldots$, $(s_i,l_i,m_i,c_i,d_i)$, $(s_{i+1},l_{i+1},m_{i+1},c_{i+1},d_{i+1})$, $\ldots$, $(s_o,l_o,m_o,c_o,d_o)$ of state transitions in $C$, \FRAM can, for example, augment an event group $E$ that corresponds to the event type $(l_i,m_i)$ with a new event that is labelled with $(l_{i+1},m_{i+1})$.
We denote by $\mathit{ir}(E,C)$ the imbalance ratio of an event group that contains both the labelled events in the event group $E$ and the augmented events from $C$.
Below, we define the fitness function $\mathit{fitacc}(D,C)$ for the accuracy objective, where $\mathit{ng}(D)$ denotes the number of event groups in $D$.
\FRAM aims at maximising the fitness $\mathit{fitacc}(D,C)$.
\begin{equation*}
    \mathit{fitacc}(D,C) = \sum\limits_{E~\mathtt{in}~D}\mathit{ir}(E,C) / \mathit{ng}(D)
\end{equation*}

\paragraph{Diversity.}
\FRAM aims at testing the controller under test using diverse sequences of control messages.
To this end, at each iteration $i$ of \FRAM, it plans to guide fuzzing in the $i{+}1$th iteration to generate sequences of control messages that are different from the sequences exercised from the first to the $i$th iterations, which are captured in the event traces $D$.
Given a candidate solution $C$, \FRAM quantifies the difference between $D$ and event traces (i.e., message sequences) that can be produced by $C$ using the normalised compression distance (NCD)~\cite{CilibrasiV2005:NCD}.
NCD measures the difference between two objects $X$ and $Y$ based on their compression, the Kolmogorov complexity~\cite{Kolmogorov1965}, and the information distance~\cite{BennettGMVZ1998}.
Precisely, $\mathit{NCD}(X,Y)$ is defined as follows:
\begin{equation*}
    \mathit{NCD}(X, Y) = \frac{Z(XY) - \min\{Z(X), Z(Y)\}}{\max\{Z(X), Z(Y)\}}
\end{equation*}
where $Z()$ is an actual compressor such as gzip~\cite{rfc1952:gzip}, $Z(X)$ and $Z(Y)$ are the compressed sizes of the objects $X$ and $Y$, and $Z(X,Y)$ is the compressed size of the concatenation of $X$ and $Y$.
Note that $\mathit{NCD}(X,Y) = 0$ indicates that the two objects are identical in terms of compressed information.
In contrast, $\mathit{NCD}(X, Y) = 1 + \epsilon$ implies that they are distinct, where $\epsilon$ is a small positive value dependent on how closely the compressor $Z$ approximates the Kolmogorov complexity.
We opt to use NCD because it is applicable for comparing two sets of event traces, wherein individual events can have different message types, fuzz operators, and field values.
Furthermore, the lengths of the event traces may differ from one another, and the two sets contain different numbers of event traces.
Hence, applying simple sequence comparison methods is not straightforward in our context.

Given the event traces $D$, in order to use NCD as the diversity fitness for a candidate solution $C$, \FRAM converts $C$ into event traces $T^C$ to be generated in the subsequent iteration.
Specifically, for each transition sequence $p$ $=$ $(s_1,l_1,m_1,c_1,d_1)$, $\ldots$, $(s_i,l_i,m_i,c_i,d_i)$, $(s_{i+1},l_{i+1},m_{i+1},c_{i+1},d_{i+1})$, $\ldots$, $(s_o,l_o,m_o,c_o,d_o)$ in $C$, the sequence $p$ is converted into an event trace $\mathit{tr}$ $=$ $(l_1,m_1,\mathit{nil})$, $\ldots$, $(l_i,m_i,\mathit{nil})$, $(l_{i+1},m_{i+1},\mathit{nil})$, $\ldots$, $(l_o,m_o,\mathit{nil})$ by excluding the source and destination states $s$ and $d$ from the transitions while preserving their message type $l$ and the fuzz operator $m$, along with their original order.
Note that, in predicted event traces, field values are set to nil (i.e., $v_i$ $=$ $\textit{nil}$) since transition sequences do not capture field values.

To quantify the degree to which a candidate solution $C$ is different from the event traces $D$, we denote by $T^C$ the predicted event traces when fuzzing is guided by $C$, and below, we define the fitness function $\mathit{fitdiv}(D, C)$ to address the diversity objective.
\FRAM aims at maximising the fitness $\mathit{fitdiv}(D,C)$.
\begin{equation*}
    \mathit{fitdiv}(D,C) = \mathit{NCD}(\mathit{D}, \mathit{D} \cup \mathit{T^C})
\end{equation*}

\subsubsection{Computational search}
\label{subsubsec:search}

\begin{algorithm}[t]
\caption{Searching best candidate traces to be used in the EFSM-guided fuzzing step, based on NSGA-II.}
\label{alg:search algorithm}
\begin{algorithmic}[1]
    \Input
    \Statex $M$: An EFSM
    \Statex $D$: A set of generated event traces
    \Statex $n_p$: size of the population and the archive
    \Statex $n_s$: size of a candidate solution
    \Statex $n_k$: number of shortest paths used during the generation of a candidate solution
    \Statex $\mu_f$: candidate solution fuzzing probability
    \Statex $\mu_c$: crossover probability
    \Statex $\mu_m$ : mutation probability

    \Output
    \Statex $C_b$: Best solution
    \Statex

    \Comment{generate the initial population}
    \State $\mathbf{P} \gets \emptyset$
    \Repeat
        \State $C \gets \Call{GenerateCandidate}{M, n_s, \mu_f, n_k}$
        \State $\mathbf{P} \gets \mathbf{P} \cup C$
    \Until{$\card{\mathbf{P}} = n_p$}
    \Comment{create an empty archive}
    \State $\mathbf{P}_\alpha \gets \emptyset$
    \Repeat
        \Comment{assess the fitness of each individual}
        \ForEach{$C \in \mathbf{P}$}
            \State $f_1(C) = \mathit{fitcov}(\mathit{states}(M), C)$
            \State $f_2(C) = \mathit{fitacc}(D, C)$
            \State $f_3(C) = \mathit{fitdiv}(D, C)$
        \EndFor
        \Comment{update the archive}
        \State $\mathbf{P}_\alpha \gets \mathbf{P}_\alpha \cup \mathbf{P}$
        \State $\Call{ComputeFrontRanks}{\mathbf{P}_\alpha}$
        \State $\Call{ComputeSparsities}{\mathbf{P}_\alpha}$
        \State $\mathbf{P}_\alpha \gets \Call{SelectArchive}{\mathbf{P}_\alpha, n_p}$
        \State $\mathit{BestFront} \gets \Call{ParetoFront}{\mathbf{P}_\alpha}$
        \Comment{create a new population}
        \State $\mathbf{P} \gets \Call{Breed}{\mathbf{P}_\alpha, n_p, \mu_c, \mu_m}$
    \Until{$\mathit{BestFront}$ is the ideal Pareto front \textbf{or} the algorithm run out of time}
    \State $C_b \gets \Call{SelectOne}{BestFront}$
    \State \Return $C_b$
\end{algorithmic}
\end{algorithm}

\FRAM employs NSGA-II (Non-Dominated Sorting Genetic Algorithm II)~\cite{DebPASM2002:NSGA-II}, which has been applied in many software engineering studies~\cite{LeeSNB22,LeeSBN2023,LeeSNBP23,CaloAAHI2020,LiXCWZXWL2019,ShinNSBZ18}, to search for a near-optimal fuzzing plan (i.e., solution $C$).
Algorithm~\ref{alg:search algorithm} describes the search process.
Briefly, the algorithm first generates an initial population $\mathbf{P}$ (lines~1-6), containing $n_p$ candidate solutions.
Subsequently, the algorithm evolves the population iteratively until finding the ideal Pareto front or the allocated time budget is exhausted (line~9-24).
At each iteration, the algorithm evaluates each candidate solution $C \in \mathbf{P}$ according to the fitness functions defined in section~\ref{subsubsec:fitness functions} (line~11-15).
The algorithm then updates the archive $\mathbf{P}_\alpha$ (lines~16-17).
It then computes the Pareto ranking of the solutions in the archive $\mathbf{P}_\alpha$, along with their associated sparsity values (line~18-19).
These sparsity values quantify the distribution of optimal solutions based on their fitness values.
These ranks and sparsities are used to select the appropriate $n_p$ solutions to be kept in the archive, as well as to identify the best Pareto front (line~20-21).
The algorithm then creates a new population $\mathbf{P}$ by breeding the solutions in the archive (lines~22-23).
After the search process (lines~9-24), the algorithm returns a selected solution (lines 25-26).
Below, we describe in detail the initial population generation, breeding, and solution selection mechanisms that are specific to \FRAM.

\begin{algorithm}[t]
\caption{Creating a candidate solution}
\label{alg:create candidate solution}
\begin{algorithmic}[1]
\Input
\Statex $\mathit{M}$: EFSM to generate a candidate solution (i.e., paths on $\mathit{M}$)
\Statex $\mathit{n}$: size of a candidate solution
\Statex $\mu$ : probability of fuzzing a message
\Statex $\mathit{k}$: number of shortest paths to generate

\Output
\Statex $C$ : candidate solution
\Statex

\State $C \gets \emptyset$
\RepeatN{n}
    \State $s \gets \Call{rand\_select\_state}{M}$
    \State $P \gets \Call{find\_k\_shortest\_paths}{M,s, k}$
    \State $p \gets \Call{rand\_select\_path}{M,P}$
    \If{$\Call{rand}{0,1} \le \mu$}
        \State $\mathit{op} \gets \Call{rand\_select\_fuzz\_operator}$
        \State $\Call{associate\_fuzz\_operator}{p,s,\mathit{op}}$
    \EndIf
    \State $C \gets C \cup \{p\}$
\EndRepeatN
\State \Return $C$
\end{algorithmic}
\end{algorithm}

\paragraph{Initial population}
Given an EFSM $M$, \FRAM generates an initial population for the search, containing $n$ candidate solutions.
Algorithm~\ref{alg:create candidate solution} describes how \FRAM creates a candidate solution at the beginning of the search process (see line~4 of Algorithm~\ref{alg:search algorithm}).
Algorithm~\ref{alg:create candidate solution} takes as input an EFSM $M$, the number $n$ of transition sequences in a candidate solution $C$, a probability $\mu$ of fuzzing a message, and the number $k$ of (different) shortest paths.
At each iteration of the repeat block (lines~2-11), the algorithm finds a transition sequence $p$ to be added into $C$, and this process is repeated $n$ times.
To find a transition sequence $p$, the algorithm first randomly selects a state $s$ in $M$ (line~3).
It then finds the $k$ shortest paths from the initial state of $M$ to the selected state $s$ using the k-shortest path algorithm~\cite{Eppstein1998:KShortest} (line~4).
\FRAM uses the $k$-shortest path algorithm to obtain different transition sequences (paths) from the initial state to $s$.
Given the fuzzing probability $\mu$, the algorithm decides whether it applies a fuzz operator or not (line~6).
If the algorithm decides to apply a fuzz operator, the algorithm randomly selects one of the five fuzz operators described in Section~\ref{subsubsec:fuzz operators} (line~7).
It then associates the transition sequence $p$ with the selected fuzz operator $m$.
This guides \FRAM in the subsequent iteration to apply $m$ when the controller reaches the selected state $s$ following the transition sequence $p$.
Since we do not know what will happen after applying $m$ in the subsequent iteration of \FRAM, it allows \FRAM to potentially discover new states that are not captured in the current EFSM $M$.

\begin{figure}[t]
    \centering
    \includegraphics[width=0.75\linewidth]{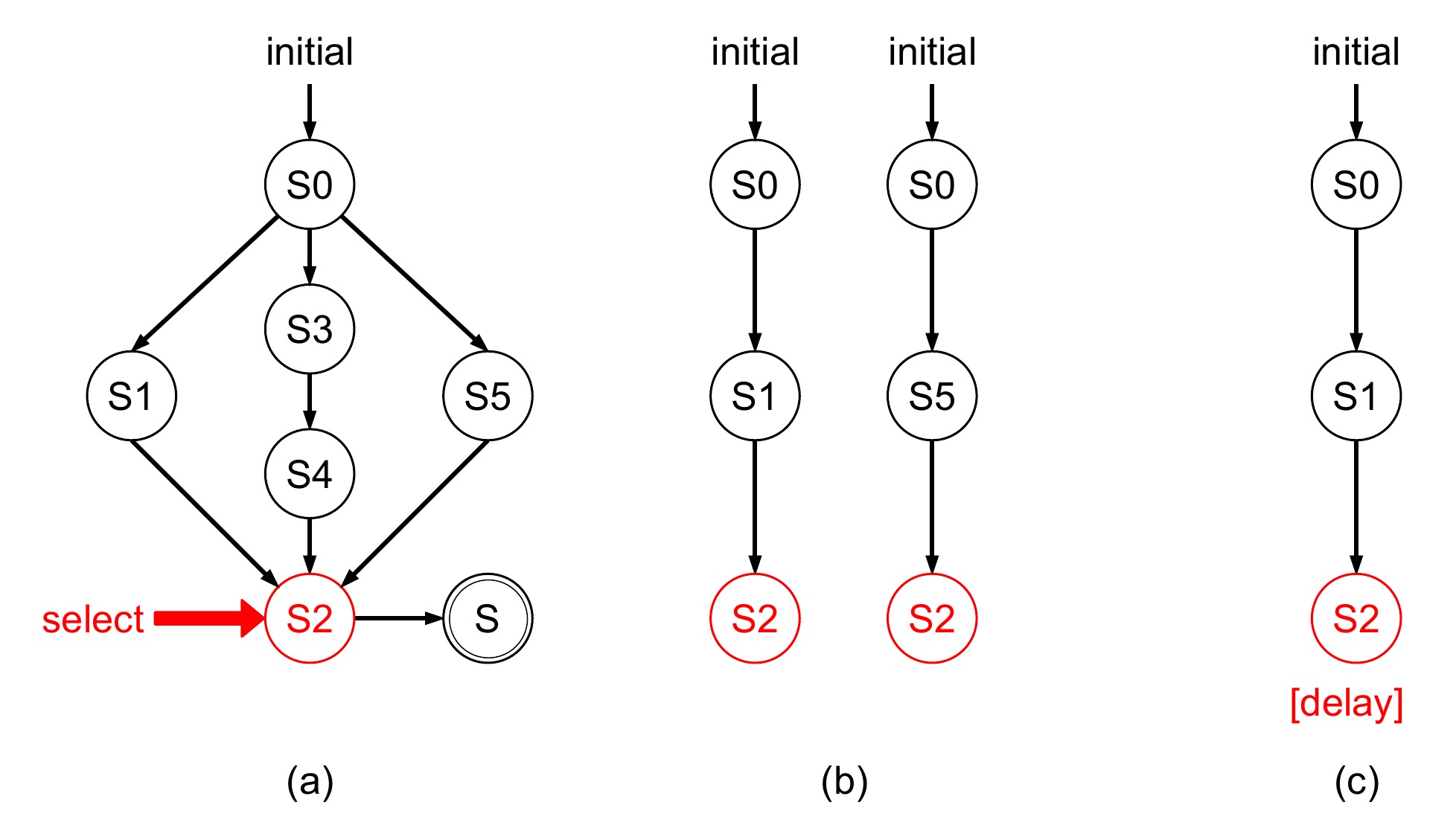}
    \caption{An example illustration of generating a candidate solution from a simple EFSM: (a)~a simple EFSM for clarity, (b)~two shortest paths from S0 to S2, and (c)~a candidate solution and its associated fuzz operator, i.e., delay.}
    \label{fig:candidate-generation}
    \Description{}
\end{figure}

For example, Figure~\ref{fig:candidate-generation} illustrates how \FRAM generates initial candidate solutions using a simple EFSM $M$ (Figure~\ref{fig:candidate-generation}~(a)) for brevity.
Given $M$, when Algorithm~\ref{alg:create candidate solution} selects state S2, it then finds two shortest paths (Figure~\ref{fig:candidate-generation}~(b)).
After that, the algorithm randomly selects a fuzz operator (e.g., delay) to apply to the selected candidate solution.

\paragraph{Breeding}
The breeding mechanism uses the following genetic operators~\cite{DebPASM2002:NSGA-II}: \emph{selection}, \emph{crossover}, and \emph{mutation} operators.
\FRAM employs the binary tournament selection and the one-point crossover~\cite{DebPASM2002:NSGA-II}.
Specifically, given two parent solutions $C^l$ and $C^r$, each containing transition sequences (paths) $\{p_1^l$, $\ldots$, $p_i^l$, $\ldots$, $p_j^l\}$ and $\{p_1^r$, $\ldots$, $p_i^r$, $\ldots$, $p_k^r\}$, respectively, the crossover operator randomly selects a crossover point $i$.
It then generates two offspring solutions by swapping transition sub-sequences separated by $i$ between the parents, resulting in $\{p_1^r$, $\ldots$, $p_i^l$, $\ldots$, $p_j^l\}$ and $\{p_1^l$, $\ldots$, $p_i^r$, $\ldots$, $p_k^r\}$.
Further, \FRAM relies on the uniform mutation operator~\cite{Talbi2009}.
Specifically, \FRAM first randomly selects a transition sequence $p$ in a candidate solution $C$.
It then replaces $p$ with a new transition sequence obtained with Algorithm~\ref{alg:create candidate solution}, setting the parameter $n$ to 1 to create a single transition sequence.

\paragraph{Selecting a near-optimal solution}
Algorithm~\ref{alg:search algorithm}, which is based on NSGA-II, outputs a set of Pareto-optimal solutions, which are equally viable with respect to the three objectives regarding coverage, accuracy, and diversity (described in Section~\ref{subsubsec:fitness functions}).
However, \FRAM requires selecting one of the solutions to guide fuzzing at the subsequent iteration.
Various methods to select a near-optimal solution in a Pareto front have been proposed in the literature, such as selecting a \emph{knee solution}~\cite{BrankeDDO2004:KneeSolutions}, or selecting a \emph{corner solution}~\cite{PanichellaKT2015} for a specific objective.
In our context, \FRAM uses a knee solution, which is often favoured in search-based software engineering studies~\cite{BrankeDDO2004:KneeSolutions, ChenLBY2018}.
This preference is due to the observation that selecting other solutions on the front to achieve a slight improvement in one objective could result in a significant deterioration in at least one other objective~\cite{BrankeDDO2004:KneeSolutions}.
Given the three objectives regarding coverage, accuracy, and diversity, \FRAM favours a candidate solution that achieves a balanced optimisation across all these objectives.

Given a selected set of candidate solutions, containing planned paths (state transitions), \FRAM finds transitions that are associated with the modification fuzz operator and a guard condition.
It then solves the guard condition using Z3 in order to apply the modification fuzz operator, ensuring the guard condition is satisfied during our EFSM-guided fuzzing (described in Section~\ref{subsec:guided fuzzing}).
For example, given a state transition $(s_i, {\uparrow}PACKET\_IN, f_k < 20 \land f_k >8, modification, s_{j})$, \FRAM solves the guard condition, such as $f_k = 10$.
When the transition is exploited during fuzzing, \FRAM modifies a PACKET\_IN message by assigning $10$ to the field $f_k$ of the message.
Note that \FRAM solves guard conditions during the (offline) planning step, rather than the (online) fuzzing step, in order to improve efficiency during fuzzing.

\subsection{EFSM-Guided Fuzzing}
\label{subsec:guided fuzzing}

After the initial fuzzing step, \FRAM uses the learning and planning outputs to guide fuzzing sequences of control messages to test the SDN controller.
This section first describes an EFSM-guided fuzzing method in \FRAM, and then illustrates the method through a running example.

\subsubsection{EFSM-guided Fuzzing Algorithm}

\begin{algorithm}[ht]
\caption{EFSM-Guided Fuzzing}
\label{alg:efsm_guided_fuzzing}
\begin{algorithmic}[1]
\Input
\Statex $M$: EFSM generated from the learning step
\Statex $C$: set of planned paths on $M$
\Output
\Statex $\mathit{seq}^\prime$ : sequences of messages after fuzzing
\Statex $C^\prime$: set of planned paths after applying one of them
\Statex

\State $\mathit{seq}^\prime \gets \langle \rangle$
\Repeat
    \State $s \gets \Call{current\_state}{M, \mathit{seq^\prime}}$
    \State $P \gets \Call{find\_applicable\_paths}{C, M, \mathit{seq^\prime}}$
    \State $\mathit{msg} \gets \Call{receive}$
    \State $\mathit{TN} \gets \Call{find\_applicable\_transitions}{s, \mathit{msg}, P}$
    \State $\mathit{op} \gets \emptyset$
    \If{$\mathit{TN} \neq \emptyset$}
        \State $\mathit{tn} \gets \Call{rand\_select}{\mathit{TN}}$
        \State $\mathit{op} \gets \Call{get\_fuzz\_operator}{tn}$
    \EndIf
    
    \If{$\mathit{op}$ is a fuzz operator}
        \State $\mathit{msg} \gets \Call{fuzz}{\mathit{msg}, \mathit{op}}$
        \State $\mathit{seq^\prime} \gets \Call{append}{\mathit{seq^\prime}, \mathit{msg}, \mathit{op}}$
    \Else
        \State $\mathit{seq^\prime} \gets \Call{append}{\mathit{seq^\prime}, \mathit{msg}}$
    \EndIf
    \State $\Call{send}{\mathit{msg}}$
\Until{the test procedure has finished}
\State $p \gets \Call{find\_used\_path}{M, C, \mathit{seq}^\prime}$
\State $C^\prime \gets C \setminus \{p\}$
\State \Return $\mathit{seq}^\prime, C^\prime$
\end{algorithmic}
\end{algorithm}

Algorithm~\ref{alg:efsm_guided_fuzzing} describes the fuzzing procedure in \FRAM once an EFSM $M$ is available, after the initial fuzzing step.
The algorithm takes as input an EFSM $M$ and a set $C$ of planning paths (i.e., sequences of state transitions) on $M$, and iterates lines~2-19 until the test procedure has finished executing.
At the beginning of each iteration, on line~3, the algorithm first identifies the current state $s$ in $M$ according to the currently observed sequence $\mathit{seq}^\prime$ of control messages.
On line~4, \FRAM then finds a set $P$ of applicable paths from the set $C$ of planning paths to guide fuzzing.
The applicable paths contain state transitions on $M$ that start from the current state $s$.
Below, Algorithm~\ref{alg:find_applicable_paths} further describes this procedure.
On line~5, the algorithm receives a control message $\mathit{msg}$ passing through the SDN control channel and then finds a set $\mathit{TN}$ of applicable transitions from $P$.
The applicable transitions start from the current state $s$, are triggered by an event type $l$ corresponding to $\mathit{msg}$, and, if there are guards, the guards hold on the field values of $\mathit{msg}$.  
On lines~7-11, if some applicable transitions are found, the algorithm randomly selects a transition $\mathit{tn}$ among the applicable transitions (line~9), and gets the fuzz operator $\mathit{op}$ of $\mathit{tn}$, if $\mathit{tn}$ has one (line~10).
On lines~12-15, if the fuzz operator $\mathit{op}$ is present, the algorithm applies it to $\mathit{msg}$ (line~13) and appends it (line~14) to the output sequence $\mathit{seq}^\prime$ along with the applied fuzz operator ($\mathit{op}$).
On lines~16-17, if no fuzz operator is present, the algorithm simply appends the originally received message $\mathit{msg}$ to the output sequence $\mathit{seq}^\prime$.
On line~18, the algorithm sends back the $\mathit{msg}$, which could be fuzzed, into the control channel.
Since, at each iteration of \FRAM, the test procedure is run multiple times, on lines~20-21, the algorithm removes a planned path that has been applied, enabling the subsequent executions of the test procedure to be fuzzed, guided only by the remaining planned paths.

\begin{algorithm}[t]
\caption{Finding Applicable Paths}
\label{alg:find_applicable_paths}
\begin{algorithmic}[1]
\Input
\Statex $C$: set of planned paths
\Statex $M$: EFSM
\Statex $\mathit{seq}$: sequence of messages
\Output
\Statex $C^\prime$: set of applicable paths
\Statex

\State $C^\prime \gets \emptyset$
\State $p^s \gets \Call{walk}{M, \mathit{seq}}$
\State $s \gets \Call{current\_state}{M, \mathit{seq^\prime}}$
\ForEach{$p \in C$, where $s$ is on $p$}    
    \State $p^\prime \gets \Call{SubPath}{p, s}$   
    \If{for all $s \in p^\prime$, $s \in p^s$, and all $s$ appear in the same order on both $p^\prime$ and $p^s$}
        \State $C^\prime \gets C^\prime \cup \{p\}$
    \EndIf
\EndFor
\State \textbf{return} $C^\prime$
\end{algorithmic}
\end{algorithm}

Algorithm~\ref{alg:find_applicable_paths} identifies a set $C^\prime$ of applicable paths on an EFSM $M$ based on a given sequence $\mathit{seq}$ of control messages and a set $C$ of planned paths on $M$.
On lines 1-3, the algorithm initialises a return set $C^\prime$ of applicable paths on $M$, converts $\mathit{seq}$ into a path $p^s$ on $M$, and identifies the current state $s$ on $M$ for the given $\mathit{seq}$.
The algorithm then examines each path $p$ in $C$ to determine whether the current state $s$ appears on $p$ (line~4) and whether the sub-path $p^\prime$ of $p$ from the start state to $s$ in $M$ is a derivative of $p^s$ that corresponds to the sequence $\mathit{seq}$ of control messages (lines~5-6).
If $p^\prime$ is a derivative of $p^s$, $p^\prime$ can be derived from $p^s$ by deleting some transitions without changing the order of the remaining transitions.
Recall from Algorithm~\ref{alg:create candidate solution} that \FRAM uses the $k$-shortest path algorithm to create planned paths.
Hence, Algorithm~\ref{alg:find_applicable_paths} checks whether planed (sub-)paths on $M$ are derivatives of the paths in $M$ that correspond to sequences of control messages.
The algorithm then returns a set $C^\prime$ of applicable paths that satisfy the conditions described above.

\subsubsection{EFSM-Guided Fuzzing Example}

\begin{figure}[t]
\captionsetup[subfigure]{position=b}
\begin{subfigure}{\columnwidth}
\centering
\includegraphics[width=0.6\columnwidth]{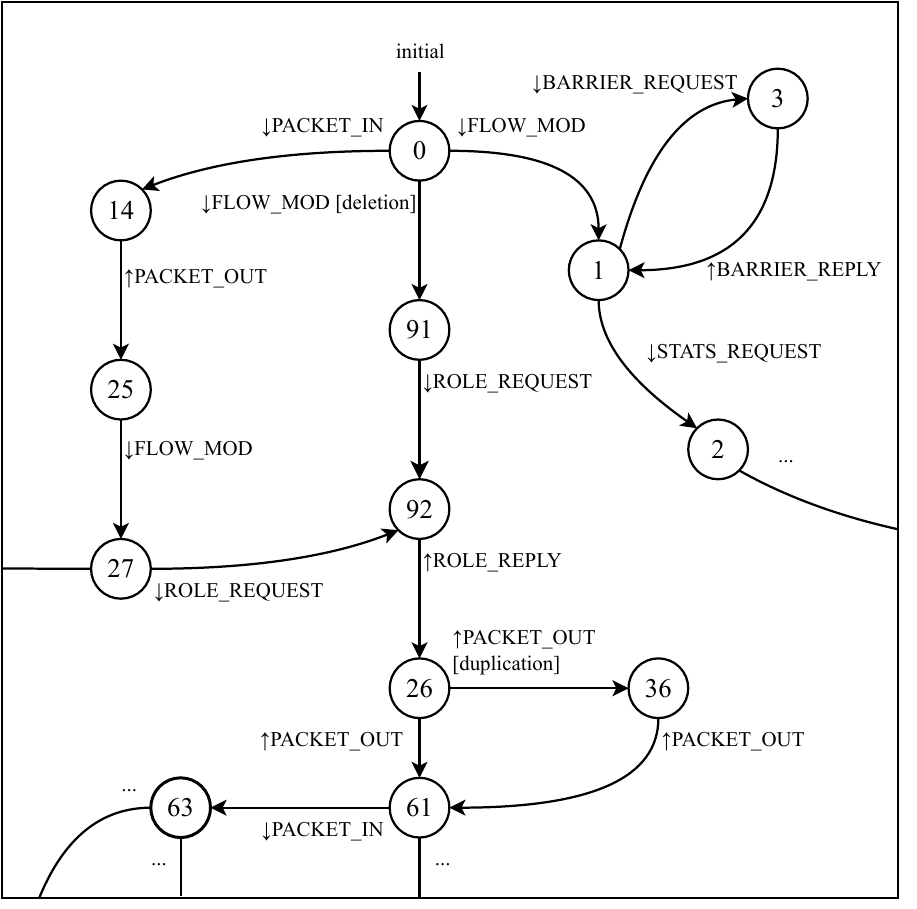}
\caption{A partial EFSM example inferred from the learning step of \FRAM.}
\label{fig:running_example_efsm}    
\end{subfigure}
\hfill\vspace{0.5em}
\begin{subfigure}{\columnwidth}
\centering
\includegraphics[width=0.8\columnwidth]{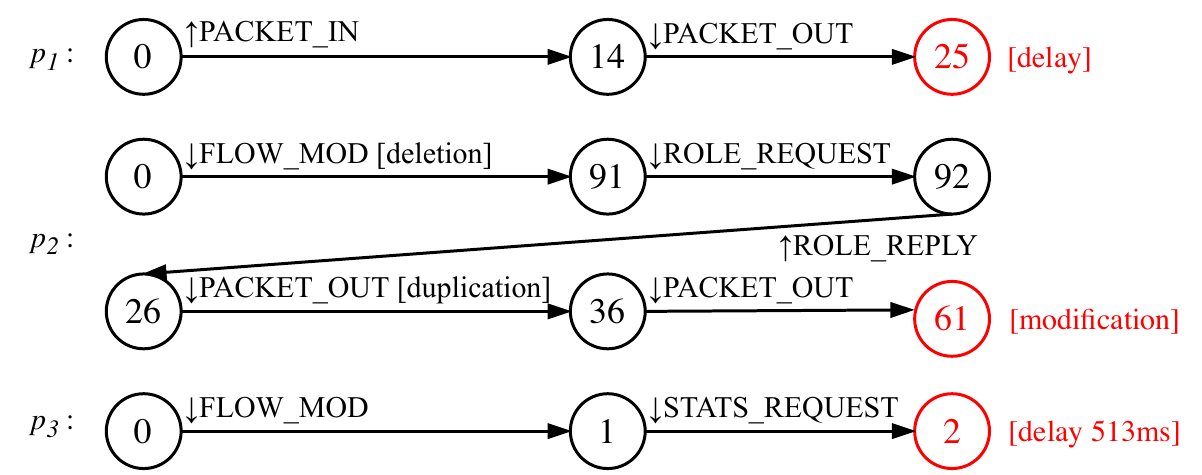}
\caption{Three planned paths to guide fuzzing, created by the planning step of \FRAM.}
\label{fig:running_example_paths}    
\end{subfigure}
\caption{Output examples of the learning and planning steps: (a)~a partial EFSM and (b)~three planned paths on the EFSM.}
\label{fig:efsm-guided fuzzing example}
\Description{}
\end{figure}

Figure~\ref{fig:efsm-guided fuzzing example} presents a part of an EFSM $M$ (Figure~\ref{fig:running_example_efsm}) inferred from the learning step and three planned paths $C$ (Figure~\ref{fig:running_example_paths}) in $M$ created by the planning step.
Given the EFSM $M$ and the planned paths $C$, when \FRAM begins executing the test procedure (e.g., ping test), Algorithm~\ref{alg:efsm_guided_fuzzing} starts with the initial state $0$ in $M$ to perform EFSM-guided fuzzing.
Then all the three planned paths, $p_1$, $p_2$, and $p_3$ shown in Figure~\ref{fig:running_example_paths}, are identified as applicable paths.
The algorithm then receives the first control message (generated by the test procedure), which, in this example, we assume to be ``$\uparrow$HELLO''.
In this case, however, there are no planned paths that contain transitions starting from state $0$ and taking the event (message) ``$\uparrow$HELLO''.
Hence, the ``$\uparrow$HELLO'' message is sent back into the control channel without any modification.
The ``$\uparrow$HELLO'' message is then appended to the output sequence $\mathit{seq^\prime}$, as follows:
\begin{equation*}
    \mathit{seq^\prime} = \langle \mathtt{{\uparrow}HELLO} \rangle
\end{equation*}

After receiving the ``$\uparrow$HELLO'' message, in this example, the algorithm receives three more control messages, as follows: ``$\downarrow$HELLO'', ``$\downarrow$FEATURES\_REQUEST'', and ``$\uparrow$FEATURES\_REPLY''.
In these cases, the EFSM $M$ remains in state $0$ since there are no transitions from state $0$ that can be taken by the three messages.
In addition, there are no applicable paths.
As a result, those messages are sent back to the control channel, and the output sequence $\mathit{seq^\prime}$ is as follows:
\begin{equation*}
    \mathit{seq^\prime} = \langle
    \mathtt{{\uparrow}HELLO},\ 
    \mathtt{{\downarrow}HELLO},\
    \mathtt{{\downarrow}FEATURES\_REQUEST},\ 
    \mathtt{{\uparrow}FEATURES\_REPLY}
    \rangle
\end{equation*}

Next, the algorithm receives the ``$\downarrow$FLOW\_MOD'' message, while the EFSM $M$ is in state $0$.
Then, the algorithm identifies paths $p_2$ and $p_3$ as applicable paths since they contain transitions starting from state $0$ and taking the event ``$\downarrow$FLOW\_MOD''.
Among the two transitions, i.e., $(0,$ ${\downarrow}FLOW\_MOD,$ $nil,$ $deletion,$ $91)$ on $p_2$ and $(0, {\downarrow}FLOW\_MOD, nil, nil, 14)$ on $p_3$, the algorithm randomly selects the second one on $p_3$.
Since no fuzz operator is associated to the transition, the algorithm simply sends the message back to the control channel, and appends the message to the output sequence $\mathit{seq^\prime}$ of control messages, as follows:
\begin{equation*}
    \mathit{seq^\prime} = \langle
    \mathtt{{\uparrow}HELLO},\ 
    \mathtt{{\downarrow}HELLO},\
    \mathtt{{\downarrow}FEATURES\_REQUEST},\ 
    \mathtt{{\uparrow}FEATURES\_REPLY},\
    \mathtt{{\downarrow}FLOW\_MOD}
    \rangle
\end{equation*}

In the subsequent iteration of the algorithm, the current state of the EFSM $M$ changes to state $1$, as there is a transition from state $0$ to $1$ that takes the ``$\downarrow$FLOW\_MOD'' event.
The algorithm then finds only $p_3$ as an applicable path since it has a transition starting from state $1$. 
If the algorithm receives the ``$\downarrow$BARRIER\_REQUEST'' message, the message is forwarded to the control channel without applying any fuzz operators, as there are no applicable transitions on $p_3$, and is appended to the output sequence $\mathit{seq^\prime}$.

After this iteration, the algorithm changes the current state of the EFSM $M$ to state $6$, since there is a transition from state $1$ to $6$ taking the ``$\downarrow$BARRIER\_REQUEST'' event.
In this case, there are no applicable paths.
If the algorithm receives the ``$\downarrow$BARRIER\_REPLY'' message, it sends the message back to the control channel without any modification and updates the output sequence $\mathit{seq^\prime}$.

Since there is a transition from state $6$ to $1$ in the EFSM $M$, in the next iteration of the algorithm, the current state is set to state $1$.
Path $p_3$ is applicable in this situation.
If the algorithm receives the ``$\downarrow$STATS\_REQUEST'' message, the corresponding transition on $p_3$ is identified by the algorithm.
However, since there is no fuzz operator associated to the transition, the algorithm sends the message back to the control channel and updates the output sequence $\mathit{seq^\prime}$.

In the next iteration of the algorithm, the current state of the EFSM $M$ is set to state $2$ by taking the transition from state $1$ to $2$ due to the ``$\downarrow$STATS\_REQUEST'' event.
In the state, path $p_3$ is applicable.
Note that, however, state $2$ is the end state of path $p_3$ and is associated with the delay fuzz operator, which holds a message for 513ms and then sends it back to the control channel.
This indicates that for any receiving message, the algorithm applies the delay fuzz operator.  
If the algorithm receives the ``$\uparrow$STATS\_REPLY'', it applies the delay fuzz operator.
From the subsequent iterations of the algorithm, no planned paths are applicable as $p_3$ has been exploited in its entirety.
Hence, the algorithm simply forwards the receiving messages to the control channel and updates the output sequence $\mathit{seq^\prime}$ until the end of the test procedure execution.
After executing the test procedure, we can obtain the following sequence of control messages, which leads to the SDN controller failing:
\begin{multline*}
    \mathit{seq^\prime} = \langle
    \mathtt{{\uparrow}HELLO},\ 
    \mathtt{{\downarrow}HELLO},\ 
    \mathtt{{\downarrow}FEATURES\_REQUEST},\\ 
    \mathtt{{\uparrow}FEATURES\_REPLY},\
    \mathtt{{\downarrow}FLOW\_MOD},\    \mathtt{{\downarrow}BARRIER\_REQUEST},\\
    \mathtt{{\uparrow}BARRIER\_REPLY},\ 
    \mathtt{{\downarrow}STATS\_REQUEST},\\ 
    \mathtt{{\uparrow}STATS\_REPLY [delay~513ms]},\ 
    \mathtt{{\downarrow}ERROR},\ 
    \mathtt{FAILURE}
    \rangle
\end{multline*}

 \section{Evaluation}
\label{sec:evaluation}

In this section, we empirically evaluate \FRAM.
Our complete evaluation package is available online~\cite{Artifacts}.

\subsection{Research Questions}
\label{subsec:rq}

\noindent\textbf{RQ1 (comparison):}
\textit{How does \FRAM compare against other state-of-the-art fuzzing techniques for SDNs?}
We investigate whether \FRAM can outperform state-of-the-art testing techniques for SDNs, including \DELTA~\cite{Lee2017:DELTA}, \BEADS~\cite{Jero2017:BEADS}, and \FUZZSDN~\cite{OllandoSB2023:FuzzSDN}. We choose these techniques as they rely on fuzzing to test SDN controllers and their implementations are available online.

\noindent\textbf{RQ2 (ablation study)}
\textit{How does the sampling technique employed by \FRAM influence its performance?}
We assess the impact of the sampling technique (defined in Algorithm~\ref{alg:sampling event traces}), which is our heuristic for sampling event traces to learn EFSMs.
Specifically, we assess the impact of the technique in terms of execution time, the accuracy of EFSMs, and the diversity and coverage of the fuzzing results.
To achieve this, we compare \FRAM with its variant \FRAMns, which does not sample event traces, and subsequently analyse the impact of the sampling algorithm.

\noindent\textbf{RQ3 (scalability):}
\textit{
Can \FRAM fuzz sequences of control messages and learn stateful failure-inducing models in practical time?
}
We investigate the correlation between \FRAM's execution time and network size.
To do so, we carry out experiments involving SDNs of different network sizes.

\subsection{Simulation Platform}
\label{subsec:simulation platform}
To conduct large-scale experiments, we employ a simulation platform that emulates the physical networks.
Specifically, we utilise Mininet~\cite{Lantz2010:Mininet} to create virtual networks of various sizes.
Mininet leverages real-world SDN switch programs, resulting in emulated networks that closely match real-world SDNs.
Hence, Mininet has been widely adopted in numerous SDN studies~\cite{Jero2017:BEADS, Lee2017:DELTA, Shin2020:SEAMS, OllandoSB2023:FuzzSDN}.

We note that \FRAM can also be applied to actual physical SDNs.
However, assessing \FRAM on actual physical networks through large-scale experiments, such as the ones reported in this article, is prohibitively expensive in terms of both cost and time.

Our experiments were conducted on 10 virtual machines, each equipped with 4 CPUs and 10GB of RAM.
Each experiment was conducted with a time budget of 5 days for ONOS and 3 days for RYU.
We note that, within this budget, the sensitivity values of the EFSMs generated by \FRAM reach their plateaus.
Due to the randomness of \FRAM, we repeated our experiments 10 times.
These experiments took approximately 60 days of concurrent execution on the 10 virtual machines.

\subsection{Study Subject}
\label{subsec:study subject}
We evaluate \FRAM by testing two open-source and actively maintained SDN controllers, ONOS~\cite{Berde2014:ONOS} and RYU~\cite{RYU}, both of which are still widely used in SDN studies~\cite{Lee2017:DELTA, Jero2017:BEADS, OllandoSB2023:FuzzSDN, ShuklaSSCZF20, LeeWKYPS20, LiWYYSWZ19, Zhang17}.
Both controllers' implementations are based on the OpenFlow SDN protocol specification.
\FRAM, which fuzzes OpenFlow control messages, is therefore capable of testing any SDN controller that adheres to the OpenFlow specification.

For our evaluation, we created five virtual networks with 1, 2, 4, 8, and 16 switches respectively.
Each network is managed by either ONOS or RYU.
In each network, the switches possess emulated physical connections with all the other switches, forming a fully connected topology.
Each switch is connected to two hosts, simulating devices that transmit and receive data, such as video and audio streams.

We note that the study subjects, comprising of 5 $\times$ 2 synthetic systems built on the five networks managed by ONOS and RYU, are representative of both existing SDN studies and real-world SDNs.
For instance, in prior SDN studies testing ONOS and RYU, \DELTA was evaluated using an SDN with two switches, \BEADS was evaluated using an SDN with three switches, and \FUZZSDN was evaluated on SDNs with 1, 3, 5, 7, and 9 switches, due to the significant computational resources required for conducting experiments with SDNs.

\subsection{Experimental Setup}
\label{subsec:experimental setup}

\noindent\textbf{EXP1.}
To answer RQ1, we conduct a comparative analysis of \FRAM with three other SDN testing tools: \FUZZSDN~\cite{OllandoSB2023:FuzzSDN}, \DELTA~\cite{Lee2017:DELTA}, and \BEADS~\cite{Jero2017:BEADS}.
\FUZZSDN is a testing framework that generates rule-based failure-inducing models and test cases.
\FUZZSDN employs a grammar-based machine learning-guided fuzzing technique, which enables it to progressively refine the generated failure-inducing models, offering interpretable models that describe the conditions leading to a failure.
\DELTA is a security framework designed for SDNs that allows engineers to automatically replicate established attack scenarios associated with SDNs and uncover new attack scenarios through fuzzing.
\DELTA accomplishes this by changing control messages, employing a fuzzing technique that randomises the control message byte stream, regardless of the OpenFlow protocol specificities.
Lastly, \BEADS is an automated attack discovery technique that relies on a range of mutation (fuzz) operators, with the aim of discovering attack scenarios.
\BEADS also fuzzes control messages but employs strategies such as message dropping, duplication, delay, and modification while adhering to the OpenFlow specification.
This allows \BEADS to generate fuzzed control messages that can pass beyond the message parsing layer of the system under test.

To compare \FRAM with these three SDN testing tools, we create three baselines: \FUZZSDNe, \DELTAe and \BEADSe.
These baselines extend \FUZZSDN, \DELTA and \BEADS respectively, to infer EFSMs, as the original testing tools do not produce EFSMs as part of their test outputs.
\FUZZSDNe (resp. \DELTAe and \BEADSe) encodes the fuzzed control messages and the test output (i.e., success and failure) as a dataset to infer EFSMs.
The baselines then use \textsc{Mint} to generate EFSMs.
Unlike \FRAM, \FUZZSDNe, \DELTAe, and \BEADSe do not leverage the generated EFSM to guide their fuzzing operations.

We use two synthetic systems, each with a single switch, controlled by either ONOS or RYU.
We leverage a test procedure (see Section~\ref{sec:approach}) that specifies a pairwise ping test~\cite{RFC1122:PING}, which has been used in many SDN studies~\cite{Lee2017:DELTA, Jero2017:BEADS, Dhawan2015:SPHINX, Canini2012:NICE, OllandoSB2023:FuzzSDN}.
This test procedure is important as it enables practitioners to verify communication between hosts, measure latency, detect packet loss, and identify routing issues.
For the failure detection mechanism, we identify spurious switch disconnections.
In our experiments, we identify switch disconnections that lead to communication breakdowns as failures. These failures cannot be localised using stack traces to pinpoint the causes of the failures.

In our comparison, we count the number of failures observed during the execution of \FRAM and the baselines.
In addition, from the final EFSMs inferred by the four tools, we measure the number of unique loop-free paths (corresponding to message sequences) that lead to failures.
This allows us to assess how many distinct failure-inducing sequences of state changes are captured in the EFSMs.
To further compare the four tools, we analyse the sensitivity of each EFSM, calculated using the formula:
\[
\text{sensitivity} = \frac{\#\text{accepted}}{\#\text{accepted} + \#\text{rejected}}
\]
where $\#\text{accepted}$ and $\#\text{rejected}$ are the number of traces accepted and rejected by the EFSM, respectively.
In our context, an EFSM with high sensitivity is desirable as it is less likely to miss possible failure-inducing sequences of control messages.
To fairly calculate sensitivity, we elected to create a dataset that maintains a balanced representation of success and failure traces across all tools, thereby reducing potential biases toward a specific tool.
To do so, we created a test dataset containing 800 fuzzing results, with an equal split of 400 success traces and 400 failure traces.
These fuzzing results were randomly sampled from separate runs of \FRAM, \FUZZSDN, \DELTA, and \BEADS, with each tool contributing 200 results, evenly divided into 100 success traces and 100 failure traces.
We note that other commonly used evaluation metrics, such as precision and F1-score, are not applicable in our context because an EFSM is built from fuzzed message sequences that are not generated during normal operations of the SDN controller under test.
Hence, our datasets lack negative traces---message sequences that should not be produced by the SDN controller---since they are derived from fuzzed message sequences, making it impossible to compute precision and F1-score.

Additionally, we measure the diversity of fuzzed message sequences obtained from the four tools using the Normalised Compression Distance (NCD) for multisets~\cite{CohenV2015:MultisetNCD}.
Recall from Section~\ref{sec:approach} that the fuzzed message sequences vary in length, message types, and message values, making the application of simple sequence comparison metrics difficult.
In our context, a high NCD value indicates that the fuzzed sequences of control messages (i.e., tests) are diverse, reducing the likelihood of redundancy or overly similar tests.

\noindent\textbf{EXP2.}
To answer RQ2, we compare \FRAM to its variant, named \FRAMns.
At each learning step, instead of using the sampling technique (see Algorithm~\ref{alg:sampling event traces}), \FRAMns uses all the collected event traces to infer an EFSM.

In this experiment, we use the same synthetic systems as those used in EXP1.
Our test procedure specifies a pairwise ping test, and our failure detection mechanism identifies unexpected communication breakdowns.
This experiment counts the number of iterations of the fuzzing, learning, and planning steps within the time budget and measures the execution time of each step.
In addition, we compare \FRAM and \FRAMns by measuring the sensitivity of the final EFSMs obtained after the time budget expires.
To ensure fair comparisons between \FRAM and \FRAMns, we created a test dataset comprising 1000 fuzzing results, evenly split into 500 success traces and 500 failure traces.
These results were obtained from separate runs of \FRAM, \FRAMns, \FUZZSDNe, \DELTAe, and \BEADSe, with each method contributing 200 results, evenly split into 100 success traces and 100 failure traces. 
Therefore, this test dataset is not biased toward either \FRAM or \FRAMns.
Furthermore, we measure the coverage and diversity degrees (defined in Section~\ref{subsubsec:fitness functions}) of the planned paths (corresponding to message sequences) obtained at the last iteration, allowing us to assess the effectiveness of the EFSMs in generating message sequences that cover diverse states.

\noindent\textbf{EXP3.}
To answer RQ3, we investigate the correlation between the resource consumption of \FRAM and the size of the five synthetic systems described in Section~\ref{subsec:study subject}, each with 1, 2, 4, 8, and 16 switches, controlled by either ONOS or RYU.
For this experiment, we use a test procedure that implements the pairwise ping test, similar to EXP1 and EXP2.
Compared to EXP1 and EXP2, the sequences of control messages produced by the test procedure in EXP3 differ significantly in terms of their lengths.
This is due to the fully connected topology in EXP3, which includes multiple switches.
Moreover, when there are more than two switches, the topology introduces switching loops~\cite{PetersonD2007:CompNet}, further increasing the number of events in a trace.
We measure the time required to configure Mininet and the SDN controller, perform the test procedure, and execute each step of \FRAM (i.e., fuzzing, learning, and planning).

\subsection{Parameter Setting}
\label{subsec:parameter tuning}

As described in Section~\ref{sec:approach}, \FRAM takes as input parameters that can be tuned to improve its efficiency and effectiveness.
For clarity and reproducibility, this section provides all the parameter values and describes how we set them.
We note that, given the extremely long execution time required for applying automated hyperparameter optimisation techniques in our context, we manually set some of the parameters as described below.

In the learning step, the parameters to be tuned are those of the sampling technique (Algorithm~\ref{alg:sampling event traces}) and \textsc{Mint}~\cite{Walkinshaw2016:MINT}.
For the sampling technique, we set the number ($n_{ts}$) of event traces to 1000, limiting the maximum size of the dataset used by \textsc{Mint}.
This configuration allowed \FRAM to generate EFSMs in practical time (approximately 100 minutes).
For \textsc{Mint}, we configured the parameter values of RIPPER~\cite{Cohen1995:FERI} as follows: three folds, a minimal weight of 2.0, and two optimisation runs as specified by the default setting in WEKA~\cite{WittenFH2016}.

In the planning step, we set the size of a candidate solution ($n_s$) to 200 in order to match the number of test procedure executions to be performed in each iteration of \FRAM.
This ensures that a candidate solution contains the 200 traces to be followed during the 200 executions of the test procedure.
We set the candidate solution fuzzing probability ($\mu_f$) to $0.5$, as we want to strike a balance between exploitation and exploration of the generated EFSM.
The crossover probability ($\mu_c$) and the mutation probability ($\mu_m$) in the planning step were set to $0.8$ and $0.02$, respectively, following published guidelines.
The size of the population and archive, $n_p$, is set to 100 and the search generates 50 populations, allowing the planning step to complete within a reasonable time (on average, 79 minutes for our ONOS study subject, and 44 minutes for our RYU study subject).

The remaining parameters were tuned using hyperparameter optimisation~\cite{WittenFH2016}, following guidelines from the literature~\cite{WittenFH2016, Hutter2019}.
We evaluated 10 different configurations of \FRAM using grid search~\cite{WittenFH2016}.
As a result of this optimisation process, we set the remaining parameters as follows: the initial fuzzing probability ($\mu$) of a message is 0.3, the minimum merging score ($k$) of \textsc{Mint} is 1, and the number ($n_k$) of shortest paths used during the generation of a candidate solution is 15.

The parameters of \FRAM used in our experiments could be further refined to improve efficiency and effectiveness.
However, the configuration we chose produced results that are satisfactory to support our findings.
As a result, we have not included additional experiments aimed at optimising these parameters in this article.

\subsection{Experiment Results}
\label{subsec:results}

To answer the research questions, we assessed the results obtained from both the ONOS and RYU subjects.
Since the findings from the ONOS results are consistent with those from the RYU results, this section presents only the ONOS results for brevity.
Note that the results for our RYU study subject are presented in Appendix~\ref{appendix:ryu results}.

\noindent\textbf{RQ1.}
\begin{figure}[t]
    \centering
    \includegraphics[width=\linewidth]{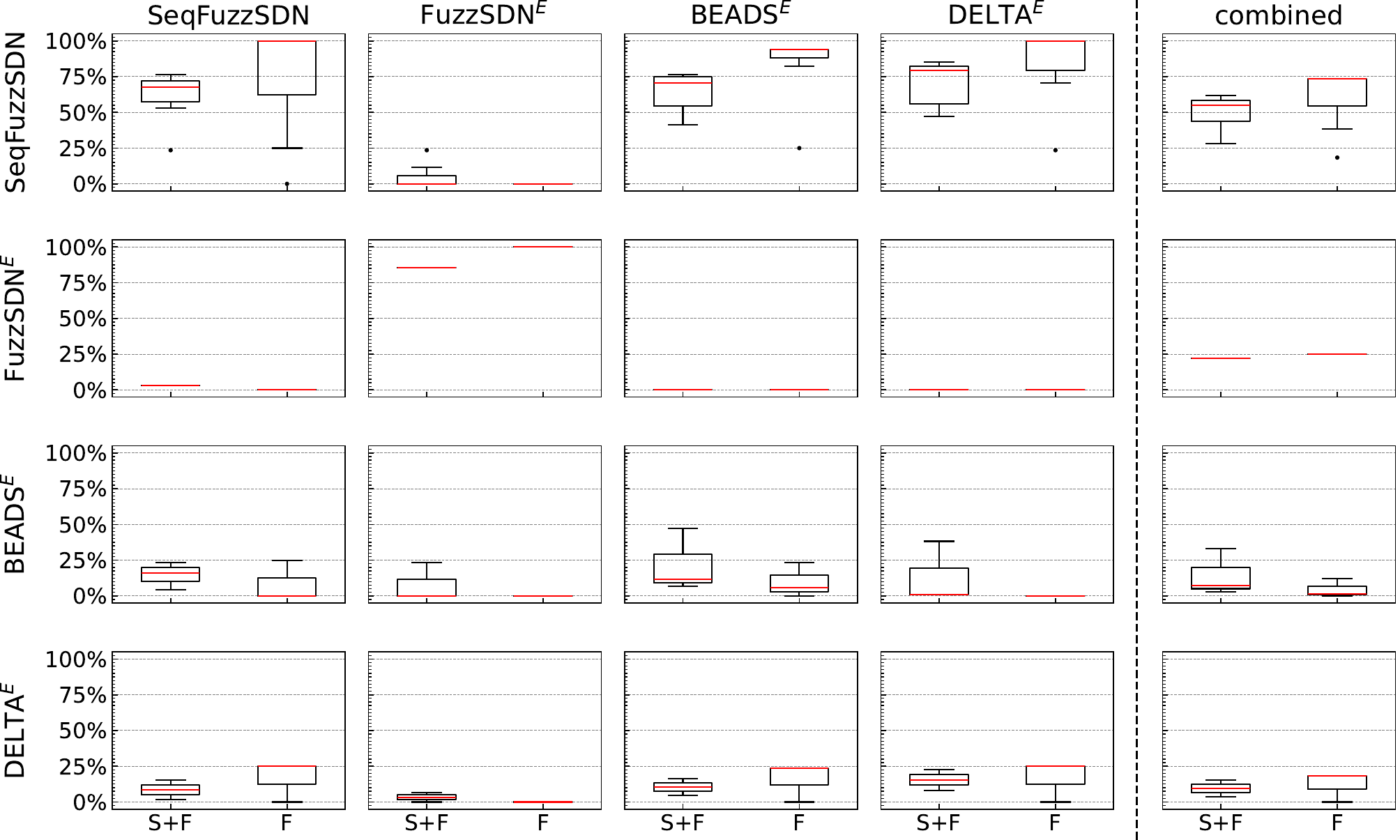}
    \caption{Comparing the sensitivity of the EFSMs generated by \FRAM, \FUZZSDNe, \BEADSe, and \DELTAe, the five plots in each row display the sensitivity of the corresponding tool.
    The first four columns represent the sensitivity of the EFSMs assessed using the test dataset containing message sequences generated by each tool.
    Sensitivity is assessed using message sequences that lead to both success and failure, denoted by (S+F), and only failure, denoted by (F).
    The last column represents the sensitivity assessed using all datasets generated by the four tools.
    The boxplots (25\%-50\%-75\%) show the distribution of sensitivity over 10 runs of each tool.}
    \label{fig:rq1-sensitivity}
    \Description{}
\end{figure}
Figure~\ref{fig:rq1-sensitivity} compares the sensitivity of the EFSMs measured using the test dataset, which contains message sequences and their test results obtained from the four tools: \FRAM, \FUZZSDNe, \BEADSe, and \DELTAe.
The last column of the first row in the figure shows that, when evaluating all the message sequences produced by these tools, on average, \FRAM achieves a sensitivity of 49.89\% on the message sequences leading to both success and failure (referred to as the combined S+F dataset) and 60.19\% on the message sequences leading only to failure (referred to as the combined F dataset).
For brevity, we refer to datasets containing message sequences generated by each tool that result in both success and failure as the [tool] S+F dataset and those that result only in failure as the [tool] F dataset.
Specifically, as shown in the first row of the figure, \FRAM achieves, on average, a sensitivity of 67.1\% on the \FRAM S+F dataset and 92.3\% on the \FRAM F dataset, 0.23\% on the \FUZZSDNe S+F dataset and 0.00\% on the \FUZZSDNe F dataset, 71.27\% on the \BEADSe S+F dataset and 86.88\% on the \BEADSe F dataset, and 73.30\% on the \DELTAe S+F dataset and 89.59\% on the \DELTAe F dataset.

For \FUZZSDNe, \BEADSe, and \DELTAe, respectively, the figure (the last column of the 2nd, 3rd, and 4th rows) shows that their EFSMs' sensitivities are, on average, 22.06\%, 14.40\%, and 9.38\% on the combined S+F dataset, and 25.00\%, 4.53\%, and 12.25\% on the combined F dataset.
Specifically, as shown in the first column of the figure, starting from the 2nd row, using the \FRAM S+F dataset (and the \FRAM F dataset), \FUZZSDNe, \BEADSe, and \DELTAe achieve, respectively, on average, sensitivities of 0.84\%, 4.78\%, and 3.68\% (and 0.0\%, 0.0\%, and 4.55\%).
Regarding the \FUZZSDNe S+F dataset (and the \FUZZSDNe F dataset), as shown in the 2nd column of the figure, \FUZZSDNe, \BEADSe, and \DELTAe achieve, respectively, on average, sensitivities of 54.62\%, 1.35\%, and 0.00\% (and 66.39\%, 0.00\%, and 0.00\%).
For the \BEADSe S+F dataset (and the \BEADSe F dataset), shown in the 3rd column, these three baselines achieve, respectively, on average, sensitivities of 0.00\%, 6.37\%, and 5.75\% (and 0.00\%, 0.00\%, and 4.28\%).
Lastly, when using the \DELTAe S+F dataset (and the \DELTAe F dataset), these baselines achieve, respectively, on average, sensitivities of 0.00\%, 5.64\%, and 8.69\% (and 0.00\%, 0.00\%, and 4.01\%).

These results show that \FRAM achieves, on average, a higher sensitivity compared to the baselines, and the differences are statistically significant.
However, note that the EFSM produced by \FRAM rejects most of the failure-inducing message sequences obtained from \FUZZSDNe, as \FRAM and \FUZZSDNe use significantly different fuzzing methods.
While \FUZZSDNe fuzzes a single message by modifying its fields' values, \FRAM fuzzes a sequence of messages using multiple fuzz operators (i.e., delay, modification, duplication, deletion, and insertion).
Consequently, the message sequences that lead to failure are significantly different between the two tools, resulting in producing very different EFSMs, which cannot accept the message sequences generated by the other tool.
Even when the same failures are triggered, the generated traces differ due to these distinct paths.
However, recall that the EFSM produced by \FUZZSDNe rejects most of the message sequences generated by \FRAM, \BEADSe, and \DELTAe, indicating that the EFSMs are specific only to \FUZZSDNe.

\begin{figure}[t]
    \centering
    \begin{subfigure}{.329\textwidth}
        \includegraphics[width=\linewidth]{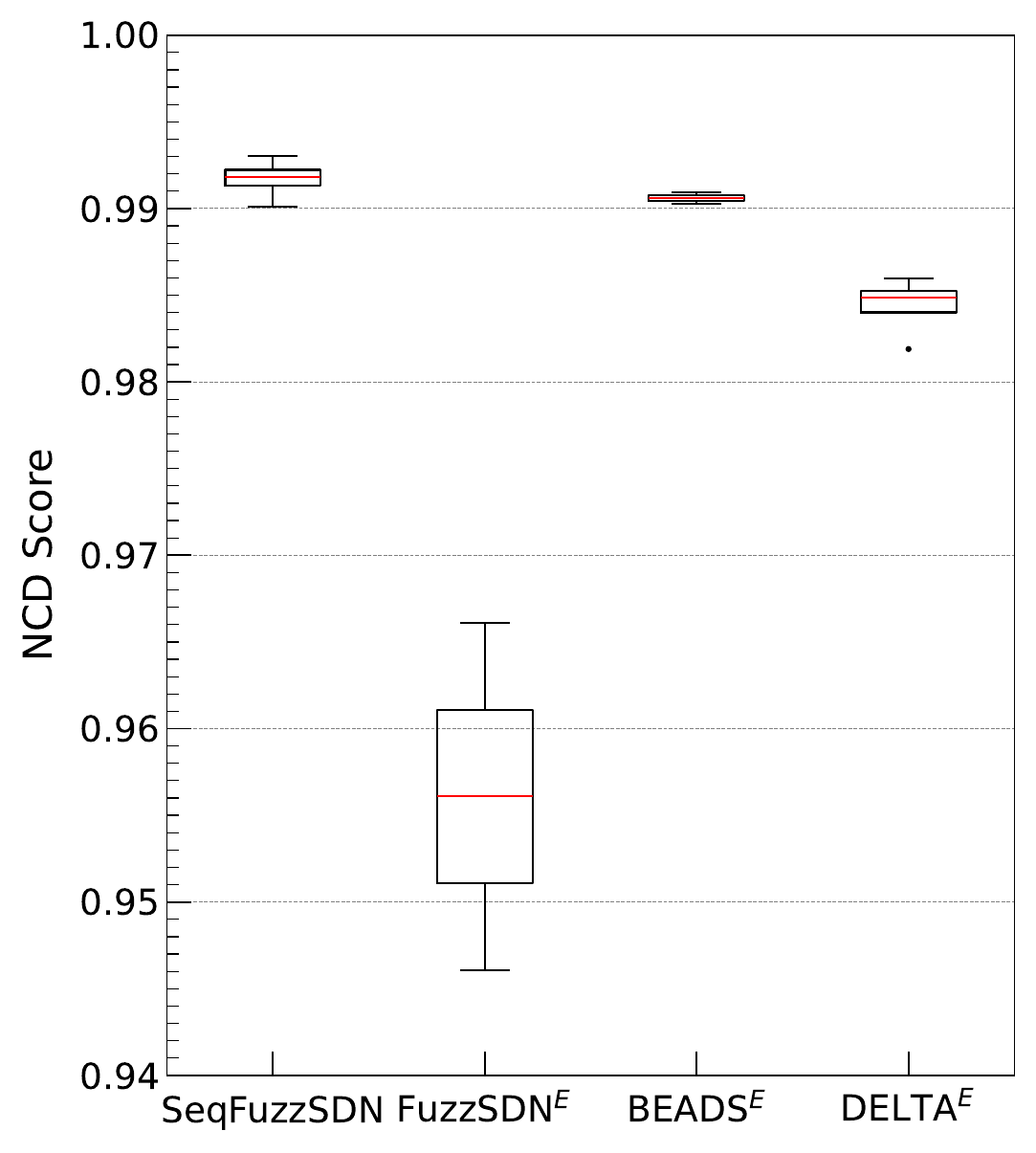}
        \caption{NCD scores}
        \label{fig:rq1-diversity}
    \end{subfigure}
    \begin{subfigure}{.329\textwidth}
        \includegraphics[width=\linewidth]{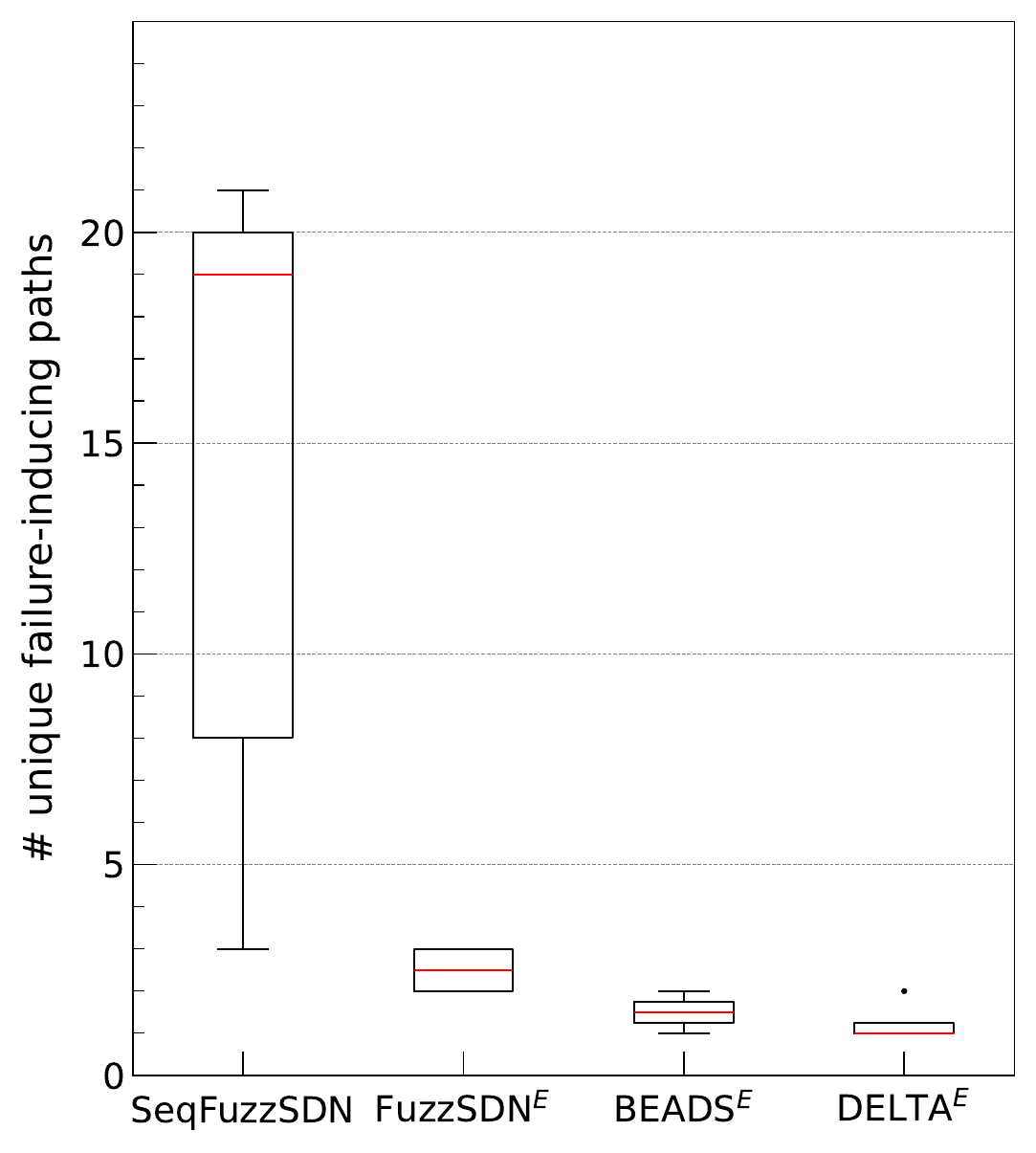}
        \caption{Number of unique failure paths}
        \label{fig:rq1-unique}
    \end{subfigure}
    \begin{subfigure}{.329\textwidth}
        \includegraphics[width=\linewidth]{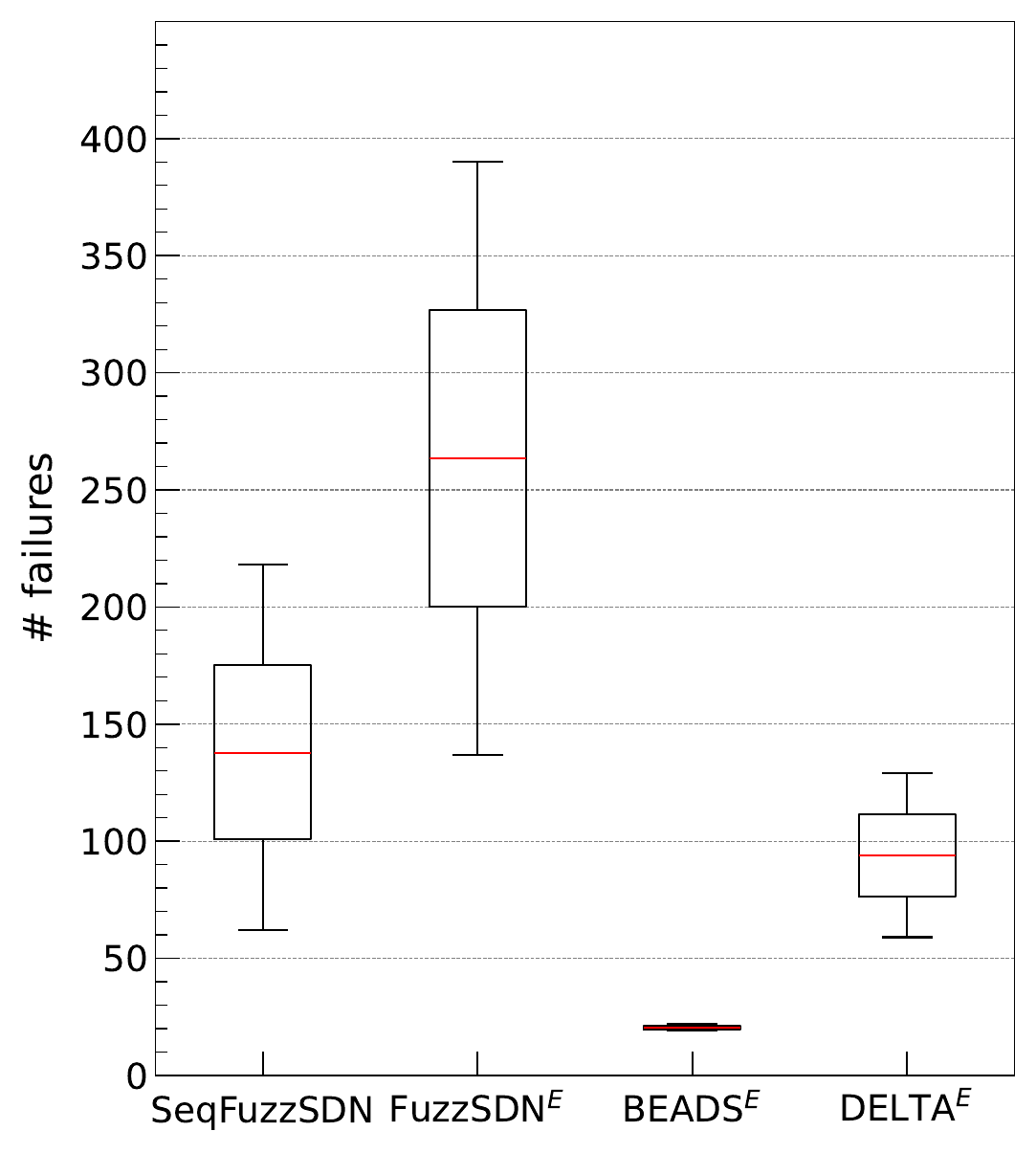}
        \caption{Number of failures}
        \label{fig:rq1-failures}
    \end{subfigure}
    \caption{Comparing (a)~the NCD scores of the message sequences, (b)~the number of unique failure-inducing paths in the EFSMs, and (c)~the number of message sequences leading to failure, all obtained from \FRAM, \FUZZSDNe, \BEADSe, and \DELTAe.
    The boxplots (25\%-50\%-75\%) show the distribution of each metric over 10 runs of each tool.}
    \label{fig:rq1-metrics}
    \Description{}
\end{figure}

Figure~\ref{fig:rq1-metrics} compares (a)~the NCD scores of the message sequences, (b)~the number of unique failure-inducing paths in the EFSMs, and (c)~the number of message sequences leading to failure, which are obtained from 10 runs of \FRAM, \FUZZSDNe, \BEADSe, and \DELTAe.
Figure~\ref{fig:rq1-diversity} shows that \FRAM achieves a higher NCD score, with an average of 0.99, compared to those of the baselines. 
Figure~\ref{fig:rq1-unique} shows that, on average, \FRAM was able to infer an EFSM containing 18 unique loop-free paths that lead to failure, which is significantly higher than the others.
From these results, we found that \FRAM generates more diverse sequences of control messages that exercise a larger number of state changes compared to the baselines.

However, Figure~\ref{fig:rq1-failures} shows that \FUZZSDNe generates a larger number of message sequences (an average of 265) leading to failure compared to the other tools, while \FRAM generates, on average, 140 message sequences leading to failure, thus outperforming \BEADSe and \DELTAe.
Even though \FUZZSDNe outperforms \FRAM in terms of number of failures, recall from Figure~\ref{fig:rq1-diversity} and Figure~\ref{fig:rq1-unique} that \FUZZSDNe generates message sequences that are less diverse and exercise significantly fewer number of state changes compared to \FRAM.
Furthermore, as described in Section~\ref{sec:approach}, \FRAM aims to generate a balanced number of message sequences that lead to success and failure, rather than focusing solely on the latter.

\begin{table}[t]
\small
\centering
\caption{Failure-inducing message sequences discovered by \FRAM in EXP1 and whether they were found by existing tools: \FUZZSDN and \BEADS.}
\label{tab:failure comparison}
\begin{tabularx}{\columnwidth}{lXcc}
\toprule
ID  & Failure-inducing message sequences & \FUZZSDN & \BEADS   \\
\midrule
ID1  & Modification of FEATURE\_REQUEST or FEATURE\_REPLY                                  & Yes & Yes \\
ID2  & Modification of *\_STATS\_REPLY of *\_STATS\_REQUEST                          & No  & Yes \\
ID3  & Removal of a non-existing flow                                                  & No  & No  \\
ID4  & Insertion of PACKET\_IN after DESC\_STATS\_REPLY                                & No  & No  \\
ID5  & Duplication of ERROR after GET\_CONFIG\_REQUEST                                 & No  & No  \\
ID6  & Deletion of GET\_CONFIG\_REPLY                                                   & No  & No  \\
ID7  & Insertion of PACKET\_IN after a flow removed                                      & No  & No  \\
ID8  & Modification of BARRIER\_REQUEST or BARRIER\_REPLY                               & Yes & Yes \\
ID9  & Modification of ECHO\_REQUEST or ECHO\_REPLY                                       & Yes & No  \\
ID10 & Insertion of PACKET\_IN after deletion of PORT\_STATS\_REQUEST                   & No  & No  \\
ID11 & Insertion of PACKET\_IN after BARRIER\_REPLY                                     & No  & No  \\
ID12 & Duplication of handshake messages                                              & No  & Yes \\
ID13 & Delay or deletion of *\_STATS\_REQUEST of *\_STATS\_REPLY                        & No  & No  \\
ID14 & Duplication of FLOW\_MOD                                             & No  & No  \\
\bottomrule
\end{tabularx}
\end{table}

In addition to the metric-based comparison described above, we reviewed the failure-inducing message sequences produced by \FRAM in EXP1 and assessed whether \FRAM could discover new failure-inducing sequences compared to existing studies.
Specifically, we manually inspected the failure-inducing sequences obtained from five runs of EXP1 (approximately 800 sequences) and categorised them based on their unique characteristics that contribute to failures into 14 cases.
Table~\ref{tab:failure comparison} presents these 14 cases of failure-inducing message sequences generated by \FRAM in EXP1, comparing them with those reported by \FUZZSDN and \BEADS.
The first column represents the class ID, the second column describes the characteristics of failure-inducing message sequences, and the third and fourth columns indicate whether or not the corresponding sequence class was identified by \FUZZSDN and \BEADS, respectively.
For example, class ID1 refers to message sequences that modify the FEATURE\_REQUEST or FEATURE\_REPLY messages.
For ID1, both \FUZZSDN and \BEADS were able to discover the failure-inducing case.
As another example, class ID10 refers to message sequences that insert the PACKET\_IN message after deleting the PORT\_STATS\_REQUEST message.
In contrast to ID1, neither \FUZZSDN nor \BEADS were able to discover this failure-inducing case.
As shown in the table, ID1 and ID8 were reported in both \FUZZSDN and \BEADS.
ID2 and ID12 were reported in \BEADS but not in \FUZZSDN.
ID9 was reported in \FUZZSDN but not in \BEADS.
IDs 3, 4, 5, 6, 7, 10, 11, 13, and 14 were not reported in either \FUZZSDN or \BEADS.
Hence, the results show that \FRAM is effective in identifying new types of failure-inducing message sequences compared to prior work.

\begin{tcolorbox}[enhanced jigsaw,left=2pt,right=2pt,top=0pt,bottom=0pt]
\emph{The answer to} \textbf{RQ1} \emph{is that} \FRAM significantly outperforms the baselines that extend \FUZZSDN, \BEADS, and \DELTA.
In particular, our experiment results indicate that \FRAM can generate more diverse sequences of control messages leading to failure than those obtained from the baselines, while also providing EFSMs that accurately capture failure-inducing message sequences.\end{tcolorbox}

\noindent\textbf{RQ2.}
\begin{figure}[t]
    \centering
    \includegraphics[width=0.5\linewidth]{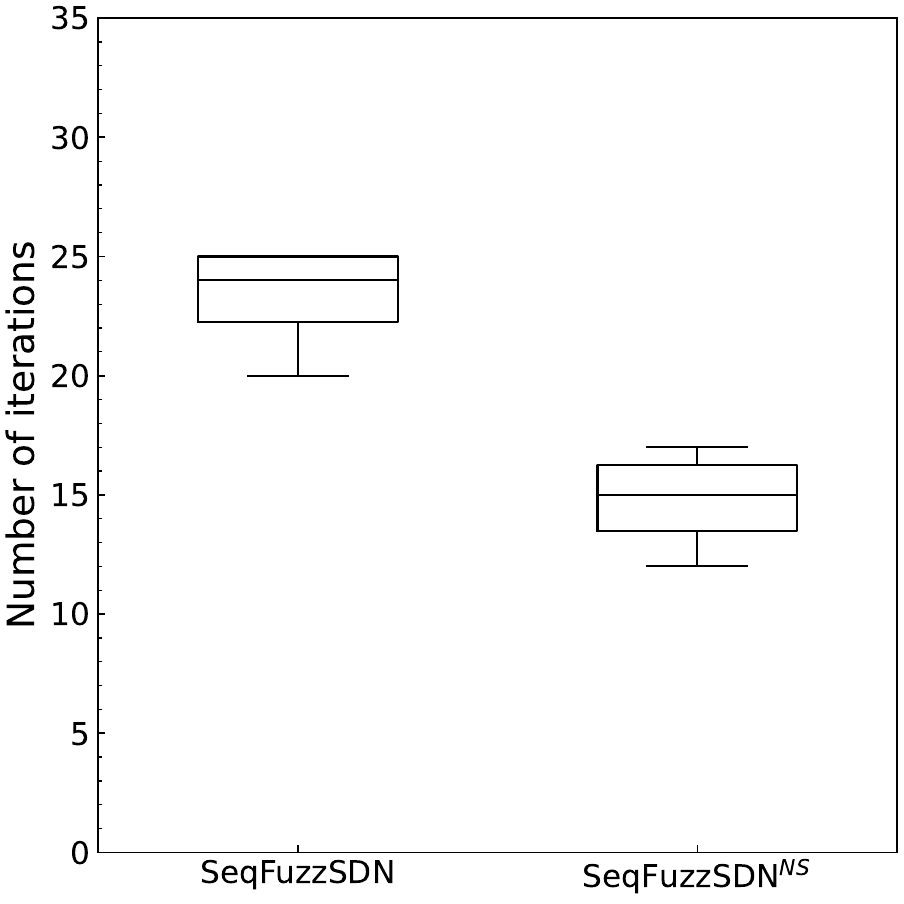}
    \caption{Comparing the number of iterations completed by \FRAM and \FRAMns within a 5-day time budget.
    The boxplots (25\%-50\%-75\%) show the distribution of iteration counts over 10 runs of each tool.} 
    \label{fig:rq2-iterations}
    \Description{}
\end{figure}
Figure~\ref{fig:rq2-iterations} presents a comparison of the number of iterations for the fuzzing, learning, and planning steps completed by \FRAM and \FRAMns within a time budget of 5 days.
The boxplots show the distributions (25\%-50\%-75\% quantiles) of the number iterations performed by \FRAM and \FRAMns, obtained from 10 runs of EXP2.
As shown in the figure, \FRAM can execute significantly more iterations than \FRAMns.
For a time budget of 5 days, \FRAM completes, on average, 25 iterations, while \FRAMns completes approximately 15 iterations.
This result indicates that the sampling technique, which caps the maximum size of the dataset for \textsc{Mint}, allows \FRAM to complete more iterations within the same time frame.
In contrast, \FRAMns, which permits the dataset to grow continuously over iterations, completes fewer iterations.
Note that each iteration of \FRAM (and \FRAMns) tests the SDN controller 200 times; hence, the sampling technique enables \FRAM to test the SDN controller, on average, 1000 times more than \FRAMns.

\begin{figure}[t]
    \centering
    \includegraphics[width=\linewidth]{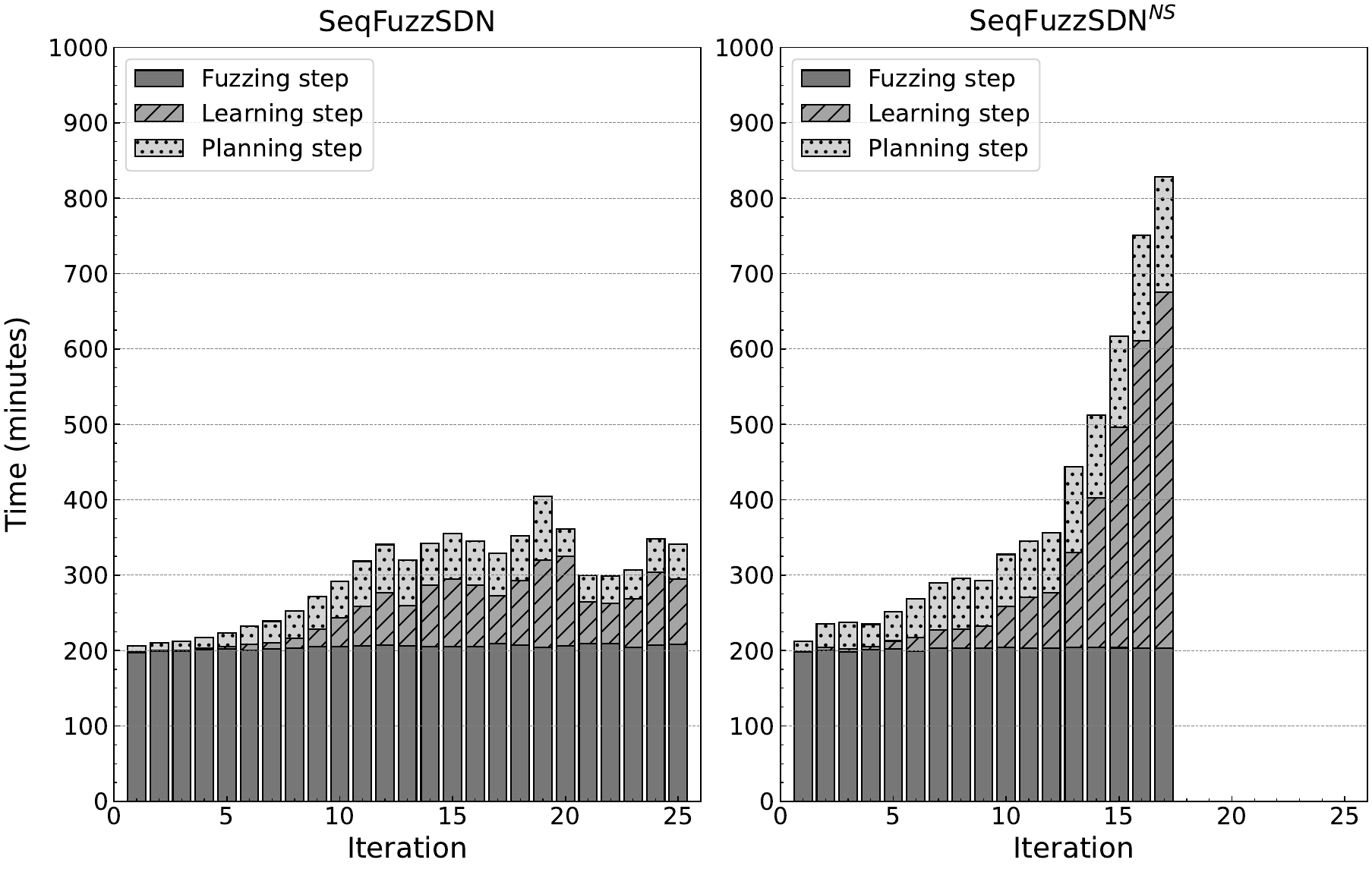}
    \caption{Comparing the execution time per iteration for the fuzzing, learning, and planning steps of \FRAM and \FRAMns within a 5-day time budget.
    The execution times shown in this figure are the average values observed over 10 runs of EXP2.} 
    \label{fig:rq2 timings}
    \Description{}
\end{figure}
In addition, Figure~\ref{fig:rq2 timings} compares \FRAM and \FRAMns with regard to the execution times per iteration for the fuzzing, learning, and planning steps over a time budget of 5 days.
The bar graph shows the average execution times taken by \FRAM and \FRAMns for the fuzzing, learning, and planning steps at each iteration, based on 10 runs of EXP2.

The results show that the fuzzing time per iteration remains constant at around 200 minutes for both \FRAM and \FRAMns, indicating that the fuzzing step is independent of the tool used.
Recall from Section~\ref{subsec:experimental setup} that EXP2 uses the pairwise ping test procedure, which is executed at each iteration during fuzzing and does not introduce variance in execution time over iterations.
For the planning step, Figure~\ref{fig:rq2 timings} shows that the planning time does not exceed 150 minutes in both \FRAM and \FRAMns.
Figure~\ref{fig:rq2 timings} also suggests that, for \FRAMns, the time required to learn an EFSM increases exponentially with each iteration due to the growing size of the dataset fed to \textsc{Mint}.
Furthermore, we observe that, on the 17th iteration of \FRAMns, the learning time reaches the 12-hour timeout limit, thus preventing \FRAMns from completing any further iterations.
This finding aligns with the literature~\cite{EmamM2018:ReHMM, WangLJB2015, ShinBB2022:PRINS}, as inferring EFSMs is a complex problem that scales poorly with larger input sizes. 
In contrast, the results for \FRAM indicate that the time required for inferring an EFSM (i.e., the learning step) remains below 115 minutes due to the application of the sampling technique.
Thus, based on the results shown in Figure~\ref{fig:rq2 timings}, we can further conclude that applying the sampling technique enables \FRAM to overcome the scalability issues associated with the complexity of learning EFSMs.

\begin{table}[t]
    \caption{Statistical significance analysis using the Wilcoxon Rank-Sum test for sensitivity, diversity, and coverage results obtained from 10 runs of EXP2.}
    \label{tbl:rq2 statistics}
    \small
    \centering
    \begin{tabularx}{\textwidth}{lYYYY}
        \toprule
         Metric            & Average (\FRAM) & Average (\FRAMns) & p-value & Statistical Significance ($\alpha = 0.05$) \\
         \midrule
         Sensitivity       & 0.542               & 0.529                 & 0.571   & Not Significant                            \\
         Diversity         & 0.9925              & 0.9920                & 0.297   & Not Significant                            \\
         Coverage   & 0.5533              & 0.6599                & 0.0124        & Significant \\
         \bottomrule
    \end{tabularx}
\end{table}
Furthermore, Table~\ref{tbl:rq2 statistics} presents the statistical test results for the distributions of sensitivity, diversity, and coverage (described in Section~\ref{sec:approach}) achieved by \FRAM and \FRAMns after 10 runs of EXP2, using the Wilcoxon Rank-Sum test~\cite{HollanderWC2015} with an $\alpha$ value of 0.05.
On average, \FRAM (resp. \FRAMns) achieves a sensitivity of 54.2\% (resp. 52.9\%), a diversity of 0.9925 (resp. 0.9920), and a coverage of 0.5533 (resp. 0.6599).
We observed that the differences in sensitivity ($p=0.14$) and diversity ($p=0.9$) are not significant, while the difference in coverage ($p=0.01$) is.
The results indicate that the use of the sampling technique does not negatively impact the sensitivity of the generated EFSMs nor the diversity of the generated message sequences.
However, the coverage achieved by \FRAM has significantly improved, suggesting that the states in the EFSM are explored more thoroughly.
One possible explanation for the improved coverage is that the increased number of iterations gives \FRAM more opportunities to refine EFSMs with respect to the coverage objective targeted at the planning step.

\begin{tcolorbox}[enhanced jigsaw,left=2pt,right=2pt,top=0pt,bottom=0pt]
\emph{The answer to} \textbf{RQ2} \emph{is that} the sampling technique introduced in \FRAM reduces its computation cost, allowing for more iterations to be performed within a given time budget.
This helps overcome scalability issues in inferring EFSMs without compromising the accuracy of the EFSMs and the diversity of the generated message sequences.
Additionally, the sampling technique significantly improves \FRAM's coverage, leading to a more thorough exploration of the search space.
\end{tcolorbox}

\begin{figure}[t]
    \centering
    \includegraphics[width=\linewidth]{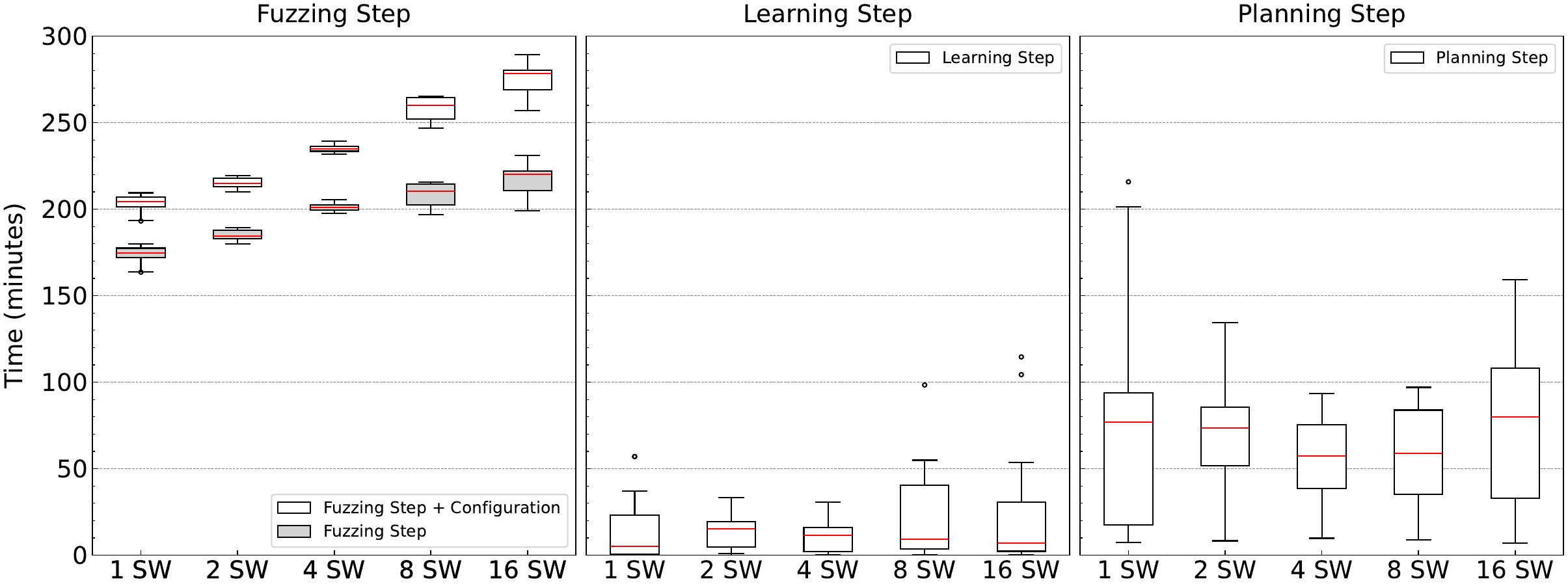}
    \caption{Boxplots (25\%-50\%-75\%) representing the distributions of time taken in minutes for the fuzzing, learning, and planning steps of \FRAM.
    This figure includes the times observed over 10 runs of \FRAM with 1, 2, 4, 8, and 16 switch configurations.} 
    \label{fig:rq3 time per phase}
    \Description{}
\end{figure}

\noindent\textbf{RQ3.}\label{par:RQ3}
Figure~\ref{fig:rq3 time per phase} presents the distributions of execution times (25\%-50\%-75\% boxplots) for the fuzzing, learning, and planning steps of \FRAM.
These execution times were measured using the five study subjects in EXP3, which consist of 1, 2, 4, 8, and 16 switches controlled by ONOS.
As shown in Figure~\ref{fig:rq3 time per phase}, the execution time taken for the fuzzing step is, on average, 203 minutes for the 1-switch configuration, 215 minutes for 2 switches, 235 minutes for 4 switches, 257 minutes for 8 switches, and 274 minutes for 16 switches.
The learning step took, on average, 15 minutes for the 1-switch configuration, 14 minutes for 2 switches, 11 minutes for 4 switches, 25 minutes for 8 switches, and 26 minutes for 16 switches. 
The planning step took, on average, 79 minutes for the 1-switch configuration, 69 minutes for 2 switches, 56 minutes for 4 switches, 56 minutes for 8 switches, and 76 minutes for 16 switches.

The results show that there is no significant difference in the times required for the learning and planning steps across the five study subjects.
However, the only time increase occurs during the fuzzing step, where test procedures are executed.
This includes the time required to configure and teardown Mininet and the SDN controller.
This increasing trend aligns with our expectations, as the execution time for a test procedure increases with its complexity.
As described in Section~\ref{subsec:experimental setup}, this is primarily due to the increasing number of messages exchanged between the switches and the controller as the number of switches and their connections grows~\cite{PetersonD2007:CompNet}.
This increase in time is independent of \FRAM, as it solely depends on the complexity of the test procedures executed.
Note that when 16 switches are fully connected, the pairwise ping test procedure produces, on average, 30.73 control messages (min: 1, max: 8178).
With 8 switches, the average number of control messages decreases to 25.19 (min: 1, max: 4863), while 4 switches produce an average of 16.16 control messages (min: 1, max: 1004).
When only 2 switches are connected, the test procedure generates, on average, 12.83 control messages (min: 1, max: 86), and with just 1 switch, the average is further reduced to 10.46 control messages (min: 1, max: 85).
The pairwise ping test procedure produces 35 unique types of control messages.
However, \FRAM halts the execution of the test procedure upon detecting a failure.
As a result, the total number of control messages recorded in our experiments could be lower than 35, depending on when failures occur.

\begin{tcolorbox}[enhanced jigsaw,left=2pt,right=2pt,top=0pt,bottom=0pt]
\emph{The answer to} \textbf{RQ3} \emph{is that} the primary factor affecting the execution time of \FRAM is its fuzzing time, which is influenced by the number of control messages generated by a test procedure.
Consequently, \FRAM is applicable to complex systems with large networks, provided that the execution time of a test procedure remains within an acceptable time budget.
\end{tcolorbox}

\subsection{Threats to Validity}
\label{subsec:threats to validity}

To address potential threats to internal validity, we compared \FRAM against three SOTA tools (\DELTA, \BEADS, and \FUZZSDN), which have been used to generate failure-inducing control messages for testing SDN controllers.
However, \DELTA, \BEADS, and \FUZZSDN do not generate failure-inducing models that consider the sequences of messages exchanged between the controller and switches.
Consequently, we extended these tools as baselines to produce EFSMs, allowing for a comparative analysis between \FRAM and these baselines.

The principal external validity threat to \FRAM is the risk that it may not be applicable to different contexts, such as other SDN systems with different switch configurations or controllers.
To address this potential threat, we conducted experiments with \FRAM against multiple SDNs and two popular SDN controllers found in the literature, namely  ONOS and RYU.
We varied our synthetic systems, which comprise five networks with 1, 2, 4, 8, and 16 switches, respectively, managed by either ONOS or RYU.

Additionally, the prototype implementation of \FRAM is compatible with OpenFlow, a widely accepted standard protocol for SDNs, which has been used in numerous SDN studies and practices~\cite{Lee2017:DELTA, Jero2017:BEADS, ShuklaSSCZF20, Canini2012:NICE, LiWYYSWZ19, LeeWKYPS20}.
As a consequence, we were able to successfully apply \FRAM to real-world SDN controllers (ONOS and RYU) and compare it to existing tools (i.e., \DELTA, \BEADS, and \FUZZSDN), considering their support for OpenFlow.
However, to further explore the applicability of our findings, it is essential to conduct additional case studies in various settings.
This includes industrial systems that use different SDN protocols and user studies that involve practitioners.

 \section{Discussion}
\label{sec:discussion}

In this section, we discuss the primary challenges faced by \FRAM and considerations  for further improvement, including testing time, testing scope, information loss due to sampling, and generalisation to other controllers.
In addition, we provide practical guidance on adapting \FRAM for other systems

\textbf{Testing time.} \FRAM executes an SDN controller for testing, which requires initialisation of the controller, network simulation, and teardown overhead.
These required operations are essential for testing the SDN controller in a realistic context.
However, further research on optimising these operations, such as reducing initialisation and teardown times and developing lightweight network simulations, is needed to enhance efficiency and scalability.

\textbf{Testing scope.} \FRAM takes as input a test procedure and a failure detection mechanism.
Hence, the testing scope is limited by these inputs.
For example, a pairwise ping test is unlikely to exercise components of the controller responsible for handling backup functionality.
Automatically exploring possible use scenarios (i.e., test procedures) and defining corresponding test oracles (i.e., failure detection mechanisms) help engineers reduce their manual efforts in creating them.
However, efficiently and effectively exploring the space of test procedures and defining oracles remain hard problems.

\textbf{Information loss.} \FRAM employs a sampling technique to use \textsc{MINT}, a model inference tool, in a scalable manner.
However, sampling sequences from all recorded message sequences inherently leads to information loss, even if our approach attempts to minimise such loss, as described in Section~\ref{subsubsec:Sampling}.
To fully address this issue, it is highly desirable to develop a scalable model inference technique capable of handling large volumes of message sequences.

\textbf{Application to other systems.} Although \FRAM was evaluated with existing fuzzing tools and SDN controllers that rely on OpenFlow, it may also be applied to other SDN controllers or even other network systems.
To facilitate such use cases, we provide practical guidance on applying and adapting \FRAM to other systems.
First, to utilise \FRAM with systems that incorporate other SDN protocols, such as Cisco OpFlex~\cite{OpFlexSpec} and ForCES~\cite{ForCESSpec}, it is necessary to modify the sniffing and injection mechanisms of \FRAM to decode and encode control messages.
However, these modifications do not affect the fuzzing, learning, and planning steps.
Therefore, we anticipate that, although such modification necessitates engineering effort to revise the sniffing and injection mechanisms, they are unlikely to impact \FRAM's efficiency and effectiveness.
Second, \FRAM may also inspire applications for testing network servers.
In such use cases, adapting \FRAM will require modifying components related to SDNs, such as parsing and modifying control messages and simulating SDN communications.
In addition, recall from Section~\ref{sec:approach} that \FRAM uses message fields to define EFSM guards and to fuzz messages.
This implies that \FRAM is applicable only to network systems where the structure of message fields is known. \section{Related Works}
\label{sec:related works}

In this section, we discuss related works in the areas of SDN testing, fuzzing, and characterising failure-inducing inputs.
Readers familiar with our previous work~\cite{OllandoSB2023:FuzzSDN} may notice significant similarities.
This is because this work builds upon, and extends our previous research.
As such, much of the foundational literature and related work remain relevant and are thus referenced here.
We believe this will provide a comprehensive context for both new readers and those familiar with our prior work.

\noindent\textbf{SDN testing.}
The study of SDN testing in the networking literature focuses on various objectives, such as detecting security vulnerabilities and attacks~\cite{NandaZDWY16, BhuniaG17, Jero2017:BEADS, Lee2017:DELTA, Zhang17, Alshanqiti19, ChicaIB20}, identifying inconsistencies among the SDN components (i.e., applications, controllers, and switches)~\cite{LiWYYSWZ19:MSAID, LeeWKYPS20, ShuklaSSCZF20}, and analysing SDN executions~\cite{Canini2012:NICE, DurairajanSB14, StoenescuDPNR18}.
In this discussion, we focus on SDN testing methods that utilise fuzzing, as they are the most relevant to our research. 
\citet{LeeWKYPS20, LeeWKNYPS22} proposed \textsc{AudiSDN}, a framework that employs a fuzzing technique to detect policy inconsistencies among SDN components (i.e., controllers and switches).
\textsc{AudiSDN} relies on the fuzzing of network policies configured by the administrators through the REST APIs of the SDN components.
To increase the probability of uncovering inconsistencies, \textsc{AudiSDN} restricts valid relationship elements by building rule dependency trees from the specification of the OpenFlow protocol.
\textsc{RE-CHECKER}, proposed by \citet{WooLKS18}, is designed to fuzz the RESTful services offered by SDN controllers.
It fuzzes an input file in JSON format, which is used by a network administrator to define network policies, such as data forwarding rules.
This process generates a large number of malformed REST messages for testing RESTful services in SDN.
\citet{DixitDS0A18} introduced \textsc{AIM-SDN} to test the implementation of the Network Management Datastore Architecture (NMDA) in SDN.
\textsc{AIM-SDN} uses random fuzzing of REST messages to test the NMDA implementation in SDN, focusing on the availability, integrity, and confidentiality of datastores.
\citet{ShuklaSSCZF20} created \textsc{PAZZ}, which is designed to identify faults in SDN switches by fuzzing data packet headers, such as IPv4 and IPv6 headers.
Finally, \citet{AlbabDHKSWTGY22} introduced \textsc{SwitchV} to verify the behaviours of SDN switches.
\textsc{SwitchV} employs fuzzing and symbolic execution to analyse the p4 models that define the behaviours of SDN switches.
In contrast to these methods, \FRAM fuzzes SDN control messages to test SDN controllers, similar to \DELTA, \BEADS, and \FUZZSDN.
Furthermore, \FRAM uses learned EFSMs to guide the fuzzing process and characterise the messages sequences that may lead to a system failure.

\noindent\textbf{Fuzzing and Stateful Testing.}
To efficiently generate effective test data, fuzzing has been widely applied in many application domains~\cite{Manes2021:ASEF}.
The research strands that most closely relate to ours are stateful fuzzing techniques~\cite{BanksCFAKV2006:SNOOZE, GasconWYAR2015:PULSAR, PhamBR2020:AFLNET, Natella22}.
Numerous research studies have explored the use of FSMs for testing complex systems.
\citet{GasconWYAR2015:PULSAR} proposed \textsc{Pulsar}, a stateful black-box fuzzing technique aimed at discovering vulnerabilities in proprietary network protocols.
Their proposed approach involves the inference of a Markov model from network traces, which are used to generate test cases using fuzzing primitives (i.e., paths in the Markov model), and finally the selection of the test cases that maximise the coverage of the protocol stack.
\citet{PhamBR2020:AFLNET} proposed \textsc{AFLNet}, a grey-box fuzzer for network protocols implementation, based on \textsc{AFL}~\cite{Zalewski:AFL}.
Their proposed technique takes a mutational approach and states feedback to guide the fuzzing of network-enabled servers.
\textsc{AFLNet} takes as input a corpus of server-client network interactions and subsequently acts as a client.
It replays modified versions of the initial message sequence sent to the server, preserving only the alterations that successfully expanded the coverage of the code or state space.
From the newly discovered message sequences, \textsc{AFLNet} uses the server’s response codes to build an FSM that describes the protocol states.
From those inferred FSMs, their approach identifies regions in the state space that have been the least explored and systematically steers the fuzzing process towards the test of such regions.
\citet{Natella22} proposed \textsc{StateAFL}, a grey-box fuzzing techniques that infers FSMs based on the in-memory states of a server, leveraging compile-time instrumentation and fuzzy hashing techniques; hence, it does not require response codes.
During the fuzzing process, \textsc{StateAFL} guides the generation of new inputs to the server based on the inferred FSMs.
It employs both byte-level and message-level fuzz operators, which do not rely on protocol specifications.
\citet{KimSMLNS2024:Ambusher} proposed \textsc{Ambusher}, a protocol-state-aware fuzzing technique for testing the \textit{``East-West''} protocol of distributed SDN controllers.
\textsc{Ambusher} takes as input a test configuration which includes the alphabet of the protocol used as well as the cluster information.
In its first phase, \textsc{Ambusher} uses a dummy network node to generate queries between the controllers, and a dummy controller to log such queries generated in the network.
In its second phase, the logged cluster queries are then used by a FSM learner to infer a FSM of the cluster's protocol.
In its third phase, \textsc{Ambusher} explores the inferred FSM to extract message sequences.
Those message sequences are then used as seeds for the fuzzing process, in which attack scenarios are generated by randomising the message sequences.
In its final phase, \textsc{Ambusher} leverages the randomised sequences to test the cluster \textit{``East-West''} interfaces.
Among these, \textsc{Ambusher} is the most relevant to \FRAM, as both take into account the SDN architecture, which differs from the server-client architecture.
Compared to \textsc{Ambusher}, \FRAM fuzzes and infers EFSMs based on sequences of control messages exchanged through the control channel of the SDN (i.e., \textit{``South''} interface).
To our knowledge, \FRAM is the first SDN testing method that focuses on the \textit{``South''} interface of SDN controllers while accounting for the statefulness of SDNs.

\noindent\textbf{Characterising Failure-Inducing Inputs.}
Recently, several research efforts have focused on identifying the input conditions that cause a system under test to fail~\cite{Gopinath2020, KampmannHSZ20}.
\citet{Gopinath2020} introduced \textsc{DDTEST}, which abstracts inputs that lead to failures.
\textsc{DDTEST} is designed to test software programs, such as JavaScript translators and command-line utilities, that accept string inputs.
It uses a derivation tree to represent how failure-inducing strings are generated.
\citet{KampmannHSZ20} developed \textsc{ALHAZEN}, which identifies the conditions under which software programs fail.
\textsc{ALHAZEN} also targets software that processes strings and uses machine learning to learn failure-inducing conditions in the form of decision trees.
In the domain of SDN systems, \citet{OllandoSB2023:FuzzSDN} introduced \FUZZSDN, a machine learning-guided Fuzzing method for testing SDN controllers.
\FUZZSDN learns an interpretable classification model that characterises conditions on a control message's fields under which the controller fails.
We note that these methods do not attempt to create a failure-inducing model for sequential data, which makes those methods not suitable for our objectives.
To our knowledge, \FRAM is the first approach that applies an EFSM-guided fuzzing approach to infer failure-inducing models, in the form of EFSMs, with a focus on SDNs.
Specifically, \FRAM tests SDN controllers by accounting for the architecture and protocols unique to SDNs, which differ from other systems (e.g., server-client systems).
Further, \FRAM tests SDN controllers without requiring any modifications or instrumentation of the controllers or their networks, enabling SDN testing in realistic operational settings.

\section{Conclusions}
\label{sec:conclusion}

We developed \FRAM, a learning-guided fuzzing method for testing stateful SDN controllers.
\FRAM uses a fuzzing strategy, guided by EFSMs, in order to (1)~efficiently explore the space of states of the SDN controller under test and (2)~infer EFSMs that characterise the sequence of messages that may make the system fail.
\FRAM implements an iterative process that fuzzes sequences of control messages, learns an EFSM, and plans how to guide the subsequent fuzzing steps by leveraging the learned EFSM.
We evaluated \FRAM on several synthetic systems controlled by two different SDN controllers.
In addition, we compared \FRAM against our extended versions of three SOTA methods for testing SDN controllers, which served as baselines in our evaluation.
Our results show that \FRAM significantly outperforms the baselines by generating effective and diverse tests (i.e., sequences of control messages), that cause the system to fail, and by producing accurate EFSMs.

In the future, we will devise a learning technique that will allow \FRAM to learn stateful models (i.e., EFSMs) incrementally, thus addressing scalability issues in inferring EFSMs.
To our knowledge, no existing solution supports incremental EFSM inference in a form applicable to \FRAM.
This poses new challenges due to the complexity of continuously updating and maintaining complex dependencies between states and transitions, without losing any of the previously learned information.
In addition, we will explore model inference techniques that can capture the asynchronous, distributed, and concurrent nature of an SDN system and apply these techniques to model SDN behaviours, utilising them to guide testing.
Since SDN components (i.e., hosts, switches, and controllers) are integrated and operate concurrently, leveraging such techniques will allow us to accurately represent interactions among components and potentially improve test effectiveness.
Further, we also aim to confirm the applicability and effectiveness of \FRAM by testing it on more SDN systems and performing user studies.

\section*{Data Availability}
\label{sec:data availability}
Our evaluation package and the \FRAM tool can be accessed online~\cite{Artifacts} to allow researchers and practitioners to (1)~reproduce our experiments and (2)~utilize and modify \FRAM.

\begin{acks}
This project has received funding from SES and the Luxembourg National Research Fund under the Industrial Partnership Block Grant (IPBG), ref. IPBG19/14016225/INSTRUCT. 
Lionel Briand was partly funded by the Science Foundation Ireland grant 13/RC/2094-2 and NSERC of Canada under the Discovery and CRC programs.
For the purpose of open access, and in fulfilment of the obligations arising from the grant agreement, the author has applied a Creative Commons Attribution 4.0 International (CC BY 4.0) license to any Author Accepted Manuscript version arising from this submission.
\end{acks}

\setcitestyle{numbers}
\bibliographystyle{ACM-Reference-Format}
\balance

\appendix
\counterwithin{figure}{section}
\counterwithin{table}{section}
\section{Additional Results for RYU Study Subject}
\label{appendix:ryu results}

\subsection{Results for RQ1}
\label{appendix:rq1 ryu}

\begin{figure}[ht]
    \centering
    \includegraphics[width=\linewidth]{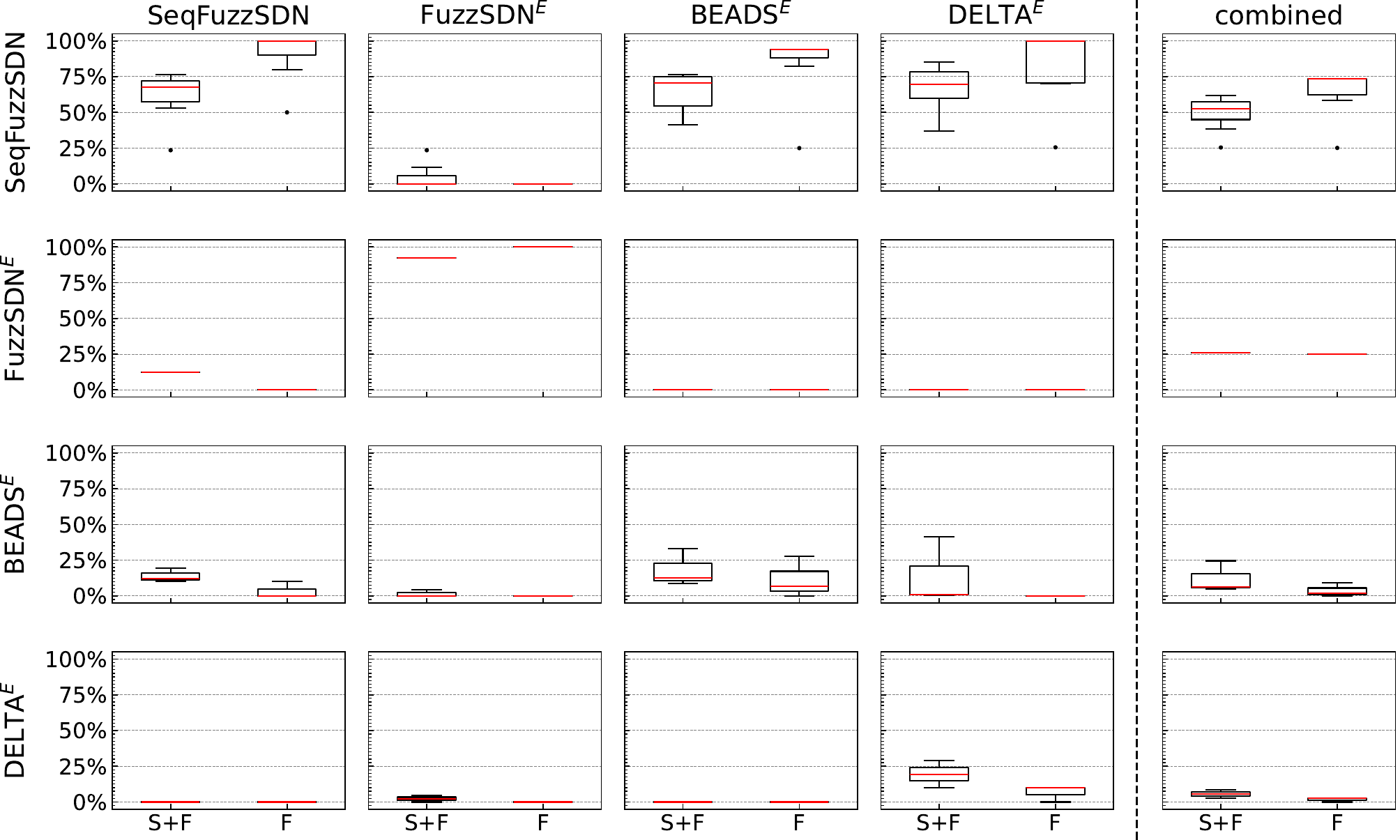}
    \caption{Comparing the sensitivity of the EFSMs generated by \FRAM, \FUZZSDNe, \BEADSe, and \DELTAe, the five plots in each row display the sensitivity of the corresponding tool.
    The first four columns represent the sensitivity of the EFSMs assessed using the test dataset containing message sequences generated by each tool.
    Sensitivity is assessed using message sequences that lead to both success and failure, denoted by (S+F), and only failure, denoted by (F).
    The last column represents the sensitivity assessed using all datasets generated by the four tools.
    The boxplots (25\%-50\%-75\%) show the distribution of sensitivity over 10 runs of each tool in EXP1 (RYU).} 
    \label{fig:rq1 sensitivity ryu}
    \Description{}
\end{figure}

In EXP1, when ONOS is replaced with RYU, Figure~\ref{fig:rq1 sensitivity ryu} corresponds to Figure~\ref{fig:rq1-sensitivity}.
Figure~\ref{fig:rq1 sensitivity ryu} shows that \FRAM achieves a sensitivity of 49.18\% on the message sequences leading to both success and failure (referred to as the combined S+F dataset) and 63.37\% on the message sequences leading only to failure (referred to as the combined F dataset).
Specifically, \FRAM achieves, on average, a sensitivity of 60.92\% on the \FRAM S+F dataset and 90.0\% on the \FRAM F dataset, 5.04\% on the \FUZZSDNe S+F dataset and 0.00\% on the \FUZZSDNe F dataset, 63.87\% on the \BEADSe S+F dataset and 82.56\% on the \BEADSe F dataset, and 66.89\% on the \DELTAe S+F dataset and 80.91\% on the \DELTAe F dataset.

Figure~\ref{fig:rq1 sensitivity ryu} also shows the average EFSM sensitivity for RYU, for \FUZZSDNe, \BEADSe, and \DELTAe, respectively, of 26.18\%, 11.96\%, and 5.43\% on the combined S+F dataset, and 25.0\%, 
3.71\%, and 1.67\% on the combined F dataset.
More specifically, using the \FRAM S+F dataset (and the \FRAM F dataset), these three baselines achieve, respectively, on average, sensitivities of 0\%, 20.41\%, and 4.87\% (and 0\%, 0.67\%, and 3.20\%).
Regarding the \FUZZSDNe S+F dataset (and the \FUZZSDNe F dataset), these three baselines achieve, respectively, on average, sensitivities of 92.29\%, 1.51\%, and 2.26\% (and 100\%, 0\%, and 0\%).
For the \BEADSe S+F dataset (and the \BEADSe F dataset), these three baselines achieve, respectively, on average, sensitivities of 0\%, 18.05\%, and 0\% (and 0\%, 11.50\%, and 0\%).
Lastly, when using the \DELTAe S+F dataset (and the \DELTAe F dataset), these baselines achieve, respectively, on average, sensitivities of 0\%, 14.24\%, and 19.44\% (and 0\%, 0\%, and 6.67\%).

\begin{figure}[t]
    \centering
    \begin{subfigure}{.329\textwidth}
        \includegraphics[width=\linewidth]{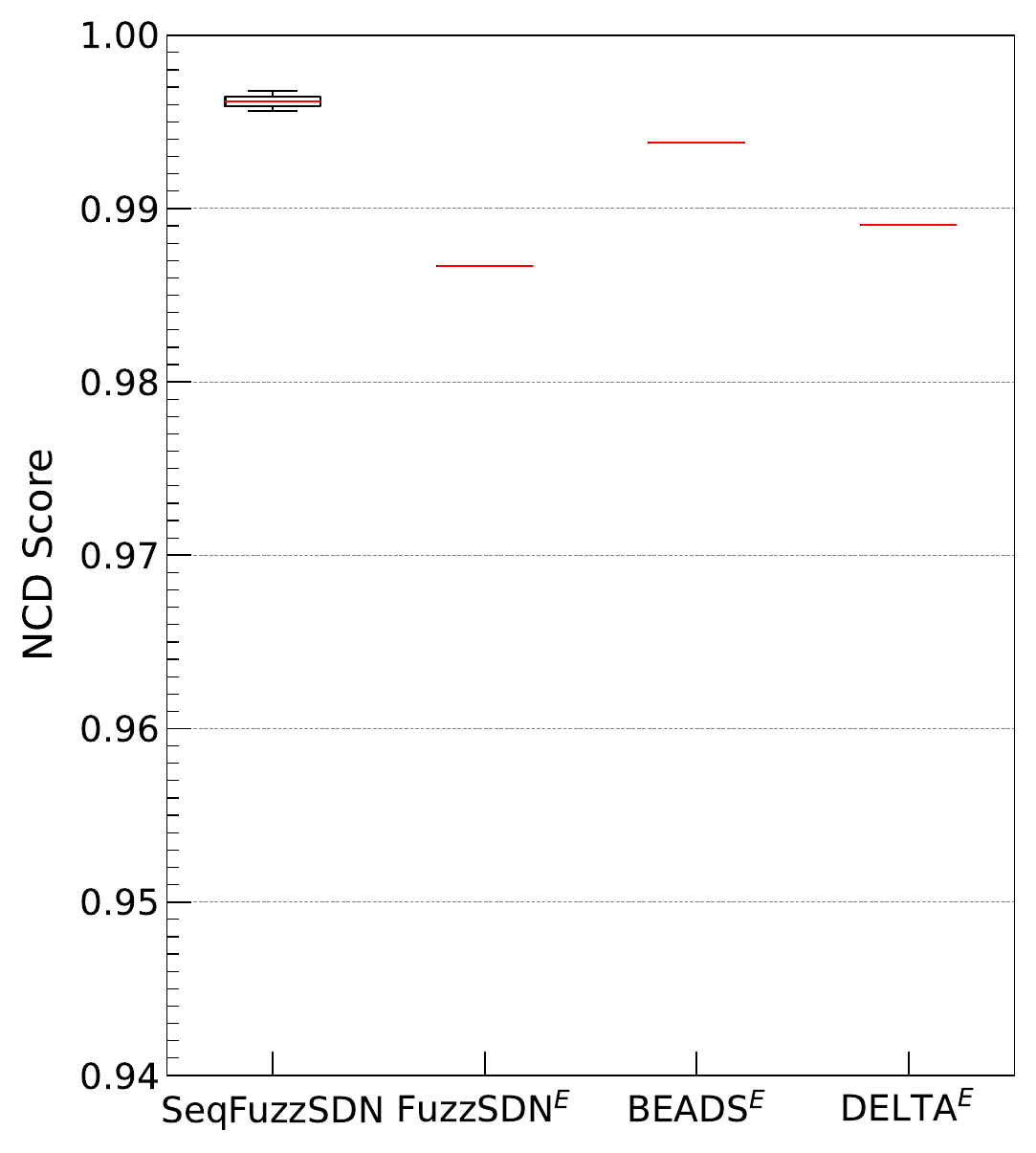}
        \caption{NCD scores}
        \label{fig:rq1-diversity ryu}
    \end{subfigure}
    \begin{subfigure}{.329\textwidth}
        \includegraphics[width=\linewidth]{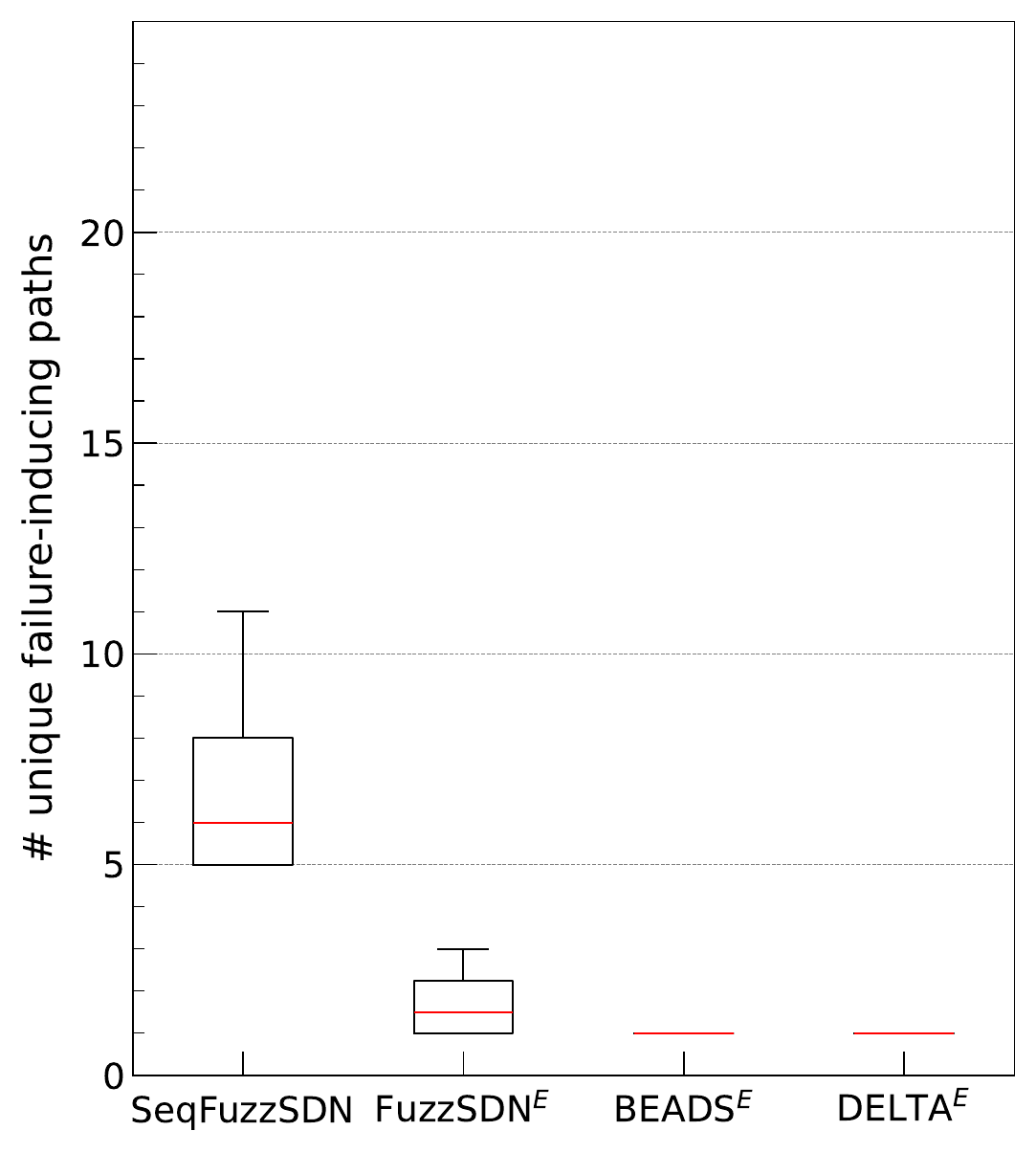}
        \caption{Number of unique failure paths}
        \label{fig:rq1-unique ryu}
    \end{subfigure}
    \begin{subfigure}{.329\textwidth}
        \includegraphics[width=\linewidth]{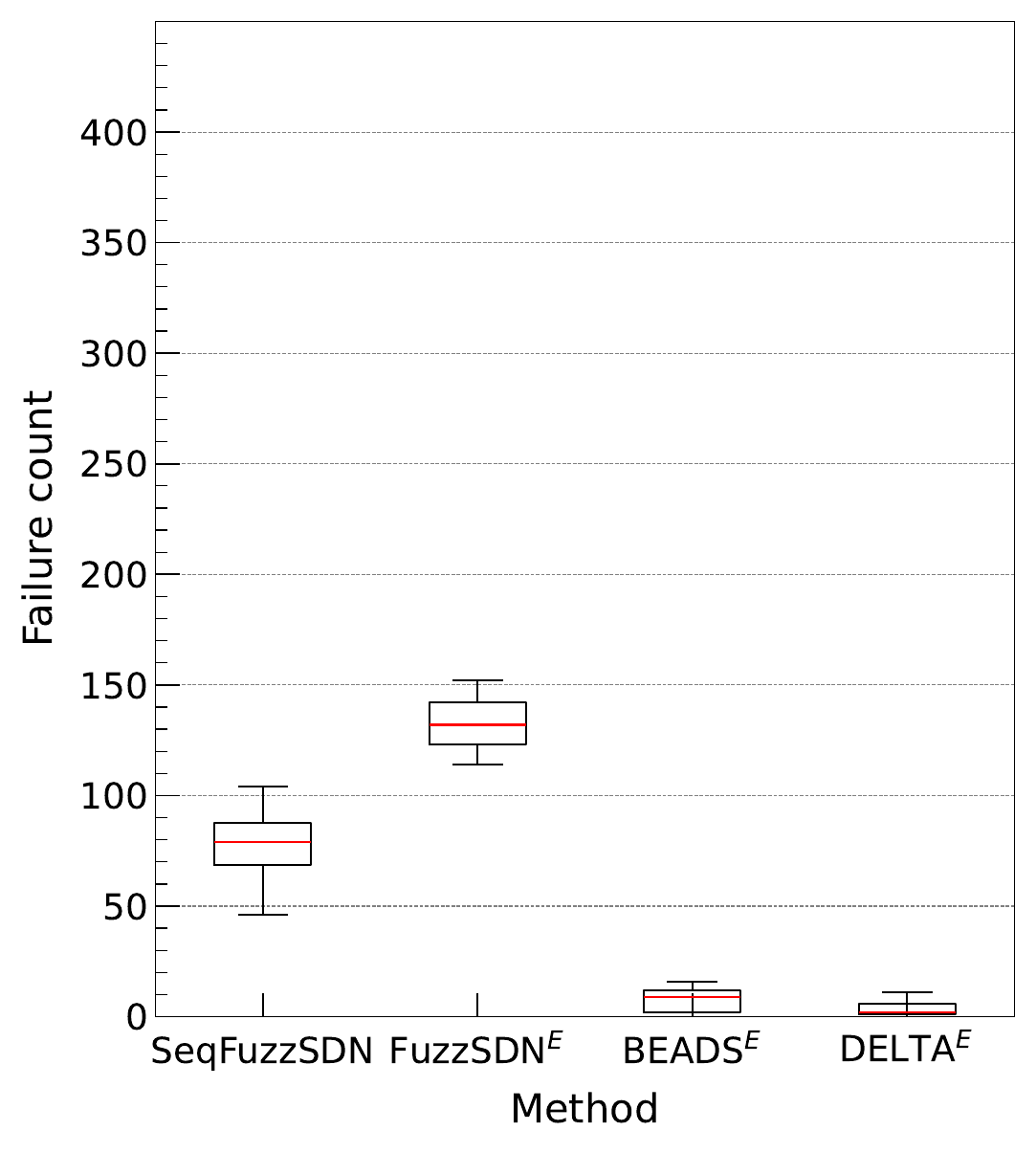}
        \caption{Number of failures}
        \label{fig:rq1-failures ryu}
    \end{subfigure}
    \caption{Comparing (a)~the NCD scores of the message sequences, (b)~the number of unique failure-inducing paths in the EFSMs, and (c)~the number of message sequences leading to failure, all obtained from \FRAM, \FUZZSDNe, \BEADSe, and \DELTAe.
    The boxplots (25\%-50\%-75\%) show the distribution of each metric over 10 runs of each tool in EXP1 (RYU).}
    \label{fig:rq1-metrics ryu}
    \Description{}
\end{figure}

Figure~\ref{fig:rq1-metrics ryu} compares (a)~the NCD scores of the message sequences, (b)~the number of unique failure-inducing paths in the EFSMs, and (c)~the number of message sequences leading to failure, which are obtained from 10 runs of \FRAM, \FUZZSDNe, \BEADSe, and \DELTAe for our RYU study subject.
Figure~\ref{fig:rq1-diversity ryu} shows that \FRAM achieves a higher NCD score, with an average of 0.997, compared to those of the baselines. 
Figure~\ref{fig:rq1-unique ryu} shows that, on average, \FRAM was able to infer an EFSM containing 6 unique loop-free paths that lead to failure, which is significantly higher than the others.
From these results, similarly to our ONOS study subject, we found that \FRAM generates more diverse sequences of control messages that exercise a larger number of state changes compared to the baselines.

However, Figure~\ref{fig:rq1-failures ryu} shows that \FUZZSDNe generates a larger number of message sequences (an average of 141) leading to failure compared to the other tools, while \FRAM generates, on average, 76 message sequences leading to failure, thus outperforming \BEADSe and \DELTAe.
As with our ONOS study subject, even though \FUZZSDNe outperforms \FRAM in terms of number of failures, recall from Figure~\ref{fig:rq1-diversity ryu} and Figure~\ref{fig:rq1-unique ryu} that \FUZZSDNe generates message sequences that are less diverse and exercise significantly fewer number of state changes compared to \FRAM.
Furthermore, as described in Section~\ref{sec:approach}, \FRAM aims to generate a balanced number of message sequences that lead to success and failure, rather than focusing solely on the latter.

\subsection{Results for RQ2}
\label{appendix:rq2 ryu}

\begin{figure}[ht]
    \centering
    \includegraphics[width=\linewidth]{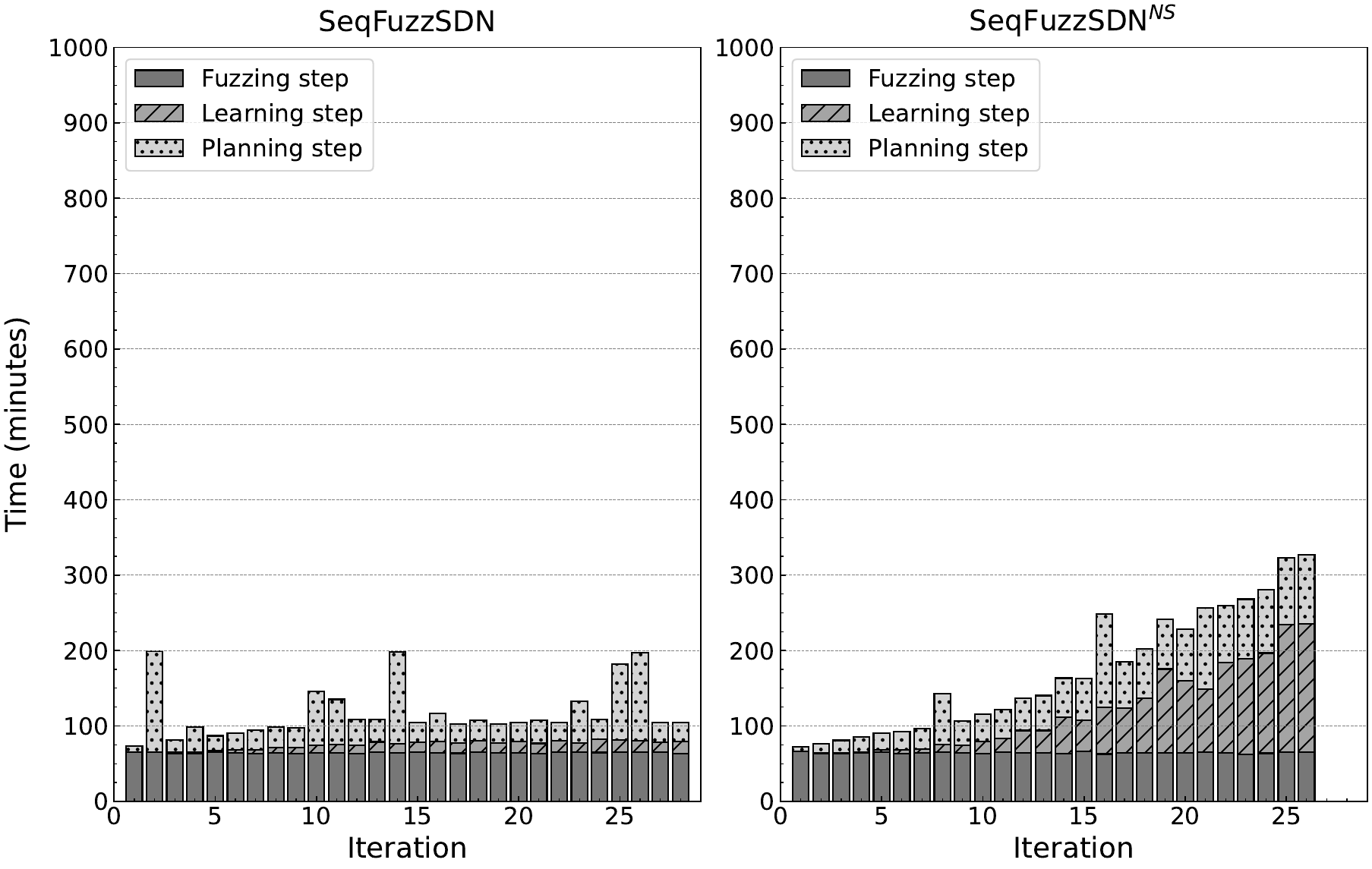}
    \caption{Comparing the execution time per iteration for the fuzzing, learning, and planning steps of \FRAM and \FRAMns within a 3-day time budget.
    The execution times shown in this figure are the average values observed over 10 runs of EXP2 (RYU).} 
    \label{fig:rq2 timings ryu}
    \Description{}
\end{figure}
Figure~\ref{fig:rq2 timings ryu} compares \FRAM and \FRAMns with regard to the execution times per iteration for the fuzzing, learning, and planning steps over a time budget of 3 days, for our RYU study subject.
Similarly to Figure~\ref{fig:rq2 timings}, the bar graph shows the average execution times taken by \FRAM and \FRAMns for the fuzzing, learning, and planning steps at each iteration, based on 10 runs of EXP2 for RYU.

The results show that the fuzzing time per iteration remains constant at around 70 minutes for both \FRAM and \FRAMns, indicating that the fuzzing step is independent of the tool used.
For the planning step, Figure~\ref{fig:rq2 timings ryu} shows that the planning time does not exceed 150 minutes in both \FRAM and \FRAMns.
Figure~\ref{fig:rq2 timings ryu} also suggests that, for \FRAMns, the time required to learn an EFSM increases significantly with each iteration due to the growing size of the dataset fed to \textsc{Mint}.
Furthermore, we observe that, as opposed to our ONOS study subject, the learning time for \FRAMns does not reach the upper learning limit of 12h, but grows from under 1 minute to above 150 minutes.
This finding aligns with the literature~\cite{EmamM2018:ReHMM, WangLJB2015, ShinBB2022:PRINS}, as inferring EFSMs is a complex problem that scales poorly with larger input sizes. 
In contrast, the results for \FRAM indicate that the time required for inferring an EFSM (i.e., the learning step) remains below 20 minutes due to the application of the sampling technique.
Thus, based on the results shown in Figure~\ref{fig:rq2 timings ryu}, we can further conclude that applying the sampling technique enables \FRAM to overcome the scalability issues associated with the complexity of learning EFSMs.

\begin{table}[ht]
    \caption{Statistical significance analysis using the Wilcoxon Rank-Sum test for sensitivity, diversity, and coverage results obtained from 10 runs of EXP2 (RYU).}
    \label{tbl:rq2 statistics ryu}
    \small
    \centering
    \begin{tabularx}{\textwidth}{lYYYY}
        \toprule
         Metric        & Average (\FRAM) & Average (\FRAMns) & p-value & Statistical Significance ($\alpha = 0.05$) \\
         \midrule
         Sensitivity & 0.553   & 0.534  & 0.385  & Not Significant \\
         Diversity   & 0.9976  & 0.9975 & 0.987  & Not Significant \\
         Coverage    & 0.5866  & 0.8248 & 0.0023 & Significant     \\
         \bottomrule
    \end{tabularx}
\end{table}
Furthermore, Table~\ref{tbl:rq2 statistics ryu} presents the statistical test results for the distributions of sensitivity, diversity, and coverage (described in Section~\ref{sec:approach}) achieved by \FRAM and \FRAMns after 10 runs of EXP2, using the Wilcoxon Rank-Sum test~\cite{HollanderWC2015} with an $\alpha$ value of 0.05, for our RYU test subject.
On average, \FRAM (resp. \FRAMns) achieves a sensitivity of 55.3\% (resp. 53.4\%), a diversity of 0.9976 (resp. 0.9975), and a coverage of 0.5866 (resp. 0.8248).
We observed that the differences in sensitivity ($p=0.18$) and diversity ($p=0.7$) are not significant, while the difference in coverage ($p=0.002$) is.
The results indicate that the use of the sampling technique does not negatively impact the sensitivity of the generated EFSMs nor the diversity of the generated message sequences, on our RYU test subject.
However, the coverage achieved by \FRAM has significantly improved, suggesting that, similarly to our ONOS test subject, the states in the EFSM are explored more thoroughly.

\subsection{Results for RQ3}
\label{appendix:rq3 ryu}

\begin{figure}[t]
    \centering
    \includegraphics[width=\linewidth]{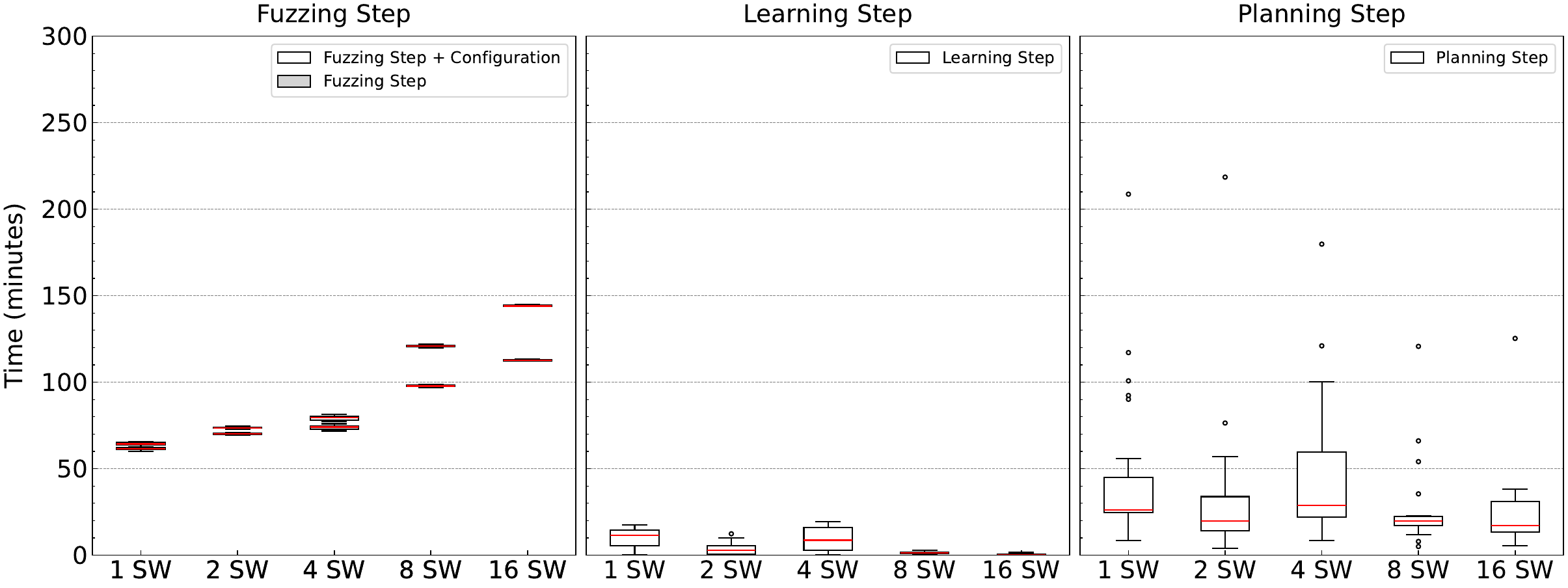}
    \caption{Boxplots (25\%-50\%-75\%) representing the distributions of time taken in minutes for the fuzzing, learning, and planning steps of \FRAM.
    This figure includes the times observed over 10 runs of \FRAM with 1, 2, 4, 8, and 16 switch configurations controlled by RYU.} 
    \label{fig:rq3 time per step ryu}
    \Description{}
\end{figure}
Figure~\ref{fig:rq3 time per step ryu} presents the distributions of execution times (25\%-50\%-75\% boxplots) for the fuzzing, learning, and planning steps of \FRAM, obtained from EXP3 (RYU).
These execution times were measured using the five study subjects in EXP3, which consist of 1, 2, 4, 8, and 16 switches controlled by RYU.
As shown in Figure~\ref{fig:rq3 time per phase}, the execution time taken for the fuzzing step is, on average, 203 minutes for the 1-switch configuration, 60 minutes for 2 switches, 70 minutes for 4 switches, 95 minutes for 8 switches, and 109 minutes for 16 switches.
The learning step took, on average, 10 minutes for the 1-switch configuration, 3 minutes for 2 switches, 9 minutes for 4 switches, 2 minutes for 8 switches, and 1 minute for 16 switches. 
The planning step took, on average, 44 minutes for the 1-switch configuration, 31 minutes for 2 switches, 47 minutes for 4 switches, 28 minutes for 8 switches, and 33 minutes for 16 switches.
These results are consistent with our findings from EXP3 (ONOS). 
\end{document}